\documentclass{myrep}
\input{psfig.sty}
\def\alt{\mathrel{\mathop
  {\hbox{\lower0.5ex\hbox{$\sim$}\kern-0.8em\lower-0.7ex\hbox{$<$}}}}}
\def\agt{\mathrel{\mathop
  {\hbox{\lower0.5ex\hbox{$\sim$}\kern-0.8em\lower-0.7ex\hbox{$>$}}}}}

\begin{document}

%%%%%%%%%%%%%%%%%%%%%%%%%%%%%%%%%%%%%%%%%%%%%%%%%%%%%%%%%%%%%%%%%%%%%%
%% Title Page %%%%%%%%%%%%%%%%%%%%%%%%%%%%%%%%%%%%%%%%%%%%%%%%%%%%%%%%
%%%%%%%%%%%%%%%%%%%%%%%%%%%%%%%%%%%%%%%%%%%%%%%%%%%%%%%%%%%%%%%%%%%%%%

\title{\vbox{\normalsize\tt
Contribution to the Proceedings of the 1997 European
School of High-Energy Physics, Menstrup near Naestved, Denmark,
25 May - 7 June 1997}\bigskip\bigskip
\rm DARK MATTER: MOTIVATION, CANDIDATES AND 
SEARCHES}

\author{G.G.~Raffelt}
\institute{Max-Planck-Institut f\"ur Physik 
(Werner-Heisenberg-Institut)\\
F\"ohringer Ring 6, 
80805 M\"unchen, Germany}

\maketitle

\begin{abstract}
  The physical nature of most of the gravitating mass in the universe
  is completely mysterious.  The astrophysical evidence for the
  presence of this dark matter and astrophysical constraints on its
  properties will be reviewed.  The most popular dark-matter
  candidates will be introduced, and current and future attempts to
  search for them directly and indirectly will be discussed.
\end{abstract}

%%%%%%%%%%%%%%%%%%%%%%%%%%%%%%%%%%%%%%%%%%%%%%%%%%%%%%%%%%%%%%%%%%%%%%
%% Section I %%%%%%%%%%%%%%%%%%%%%%%%%%%%%%%%%%%%%%%%%%%%%%%%%%%%%%%%%
%%%%%%%%%%%%%%%%%%%%%%%%%%%%%%%%%%%%%%%%%%%%%%%%%%%%%%%%%%%%%%%%%%%%%%

\section{INTRODUCTION}
\label{sec:introduction}

The question of what makes up the mass density of the universe is
practically as old as extragalactic astronomy which began with the
recognition that nebulae such as M31 in Andromeda are actually
galaxies like our own. Some of them appear in gravitationally bound
clusters. From the Doppler shifts of the spectral lines of the
galaxies in the Coma cluster, Zwicky derived in 1933 their velocity
dispersion and could thus estimate the cluster mass with the help of
the virial theorem~\cite{Zwicky}.  He concluded that the Coma cluster
contained far more dark than luminous matter when he translated the
luminosity of the galaxies into a corresponding mass.  Since then
evidence has mounted that on galactic scales and above the mass
density associated with luminous matter (stars, hydrogen clouds, x-ray
gas in clusters,~etc.)  cannot account for the observed dynamics on
those scales~\cite{Trimble,Tremaine,Boerner,KolbTurner}. In the mid
1970s it had become clear that dark matter was an unavoidable
reality~\cite{Ostriker} and by the mid 1980s the view had become
canonical that the universe is dominated by an unknown form of matter
or by an unfamiliar class of dark astrophysical
objects~\cite{Kormendy}. Besides the origin of cosmic rays and
$\gamma$-ray bursts (two major unsolved puzzles) the physical nature
of dark matter is no doubt the most stunning astrophysical mystery.

A popular hypothesis for the solution of this problem originated in a
seminal paper by Cowsik and McClelland in 1973 where they speculated
that the dark matter of galaxy clusters could consist of neutrinos if
these weakly interacting particles had a mass of a
few~eV~\cite{Cowsik73}.  About ten years earlier the cosmic microwave
background (CMB) radiation had been detected and had almost overnight
propelled the big-bang cosmogony from an obscure hypothesis to the
standard theory of the early universe. If the world originated from a
hot phase of thermal equilibrium, then all possible particles or forms
of radiation must have been produced in amounts which are easily
calculable relative to the density of microwave photons, leading to
the prediction of a ``cosmic neutrino sea'' in analogy to the
CMB. This had allowed Gershtein and Zeldovich in 1966 to derive a
stringent limit on the $\nu_\mu$ mass~\cite{Gershtein,Cowsik72}, a
second neutrino flavor which had been discovered in 1962.

\begin{figure}[b]
\centerline{\psfig{figure=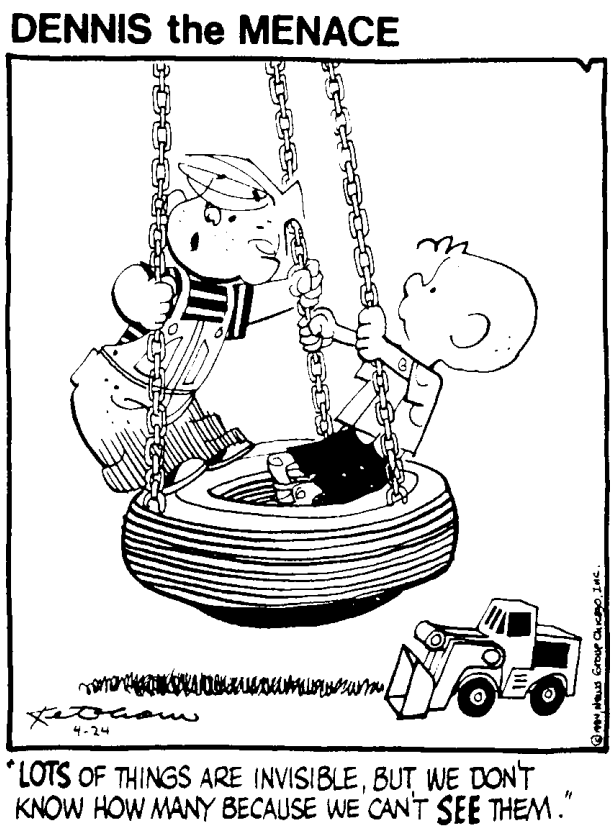,width=8.5 cm}}
\caption{DENNIS THE MENACE $^{\hbox{(R)}}$ used by permission of
Hank Ketcham and (c) by North America Syndicate.
\label{fig:dennis}}
\end{figure}

A well-known phase-space constraint on how many massive neutrinos can
be packed into a galaxy leads to a lower limit of about $20{-}30\,\rm
eV$ if they are supposed to be the dark matter in these
systems~\cite{TremaineGunn}.  This ``Tremaine-Gunn limit'' is barely
compatible with the {\em upper\/} limit of about $40\,\rm eV$ from the
overall cosmic mass density. Therefore, neutrinos certainly cannot be
the dark matter on the smallest scales where its existence is
established, most notably in dwarf galaxies. In addition, modern
theories of the formation of galaxies and larger cosmic structures
reveal that particles which stay relativistic for a long time in the
expanding universe (``hot dark matter'') prevent the formation of
small-scale structure.  Thus, even if there were enough phase space
for $40\,\rm eV$ neutrinos to be the galactic dark matter, one could
not explain how these collisionless particles would have been able to
cluster on these scales.

The alternative is ``cold dark matter,'' particles which became
nonrelativistic early. While this hypothesis works well from the
structure-formation perspective, it implies the existence of
completely new particles or else primordial black holes.  Assuming the
existence of stable weakly interacting massive particles (WIMPs) one
can predict their cosmic abundance from their mass and annihilation
cross section alone. If their interaction strength is roughly given by
Fermi's constant, then they would need a mass in the $10\,\rm GeV$
range to be the dark matter of the universe. While in the 1980s one
often discussed generic WIMPs as dark-matter candidates, the attention
today has focussed almost entirely on supersymmetric extensions of the
standard model which predict the existence of the requisite particle
in the form of a ``neutralino.'' The only other cold dark matter
candidate which is seriously discussed today are axions which are very
weakly interacting pseudoscalar bosons.

Meanwhile it is not obvious that the simplest cold dark matter
cosmologies are complete. It may be that structure formation requires
several different components, for example a certain fraction of
neutrinos plus a dominating share of neutralinos or axions (``hot plus
cold dark matter''). In addition, there may be a homogeneous mass
density in the form of vacuum energy which would play the role of a
cosmological constant. The nature of dark matter may be 
quite diverse!

The most exciting development of the 1990s is the emergence of a great
variety of real experimental projects to search for all of the
well-motivated candidates in our own galaxy. The microlensing search
for dark stars has actually produced first candidates (``MACHOs'')
which are, however, difficult to interpret.  Direct and indirect
search experiments for WIMP and axion dark matter in the galaxy have
reached a sensitivity where they begin to have a realistic chance of
finding these elusive particles. In addition, the upcoming CMB
satellites will be able to measure temperature fluctuations on very
small angular scales, allowing for a precision determination of
various cosmological parameters, notably the exact abundance of
baryonic and nonbaryonic matter. One would expect these measurements
to remove any lingering doubt about the reality of nonbaryonic dark
matter.

In these lectures I will review the astrophysical motivation for dark
matter and discuss the arguments which reveal that it is probably not
purely baryonic, and not purely in the form of massive neutrinos. I
will then proceed to discuss various candidates (dark stars,
neutrinos, WIMPs, axions) and the current attempts to search for them
by astronomical, neutrino-astronomical, and laboratory methods.

%%%%%%%%%%%%%%%%%%%%%%%%%%%%%%%%%%%%%%%%%%%%%%%%%%%%%%%%%%%%%%%%%%%%%%
%% Section II %%%%%%%%%%%%%%%%%%%%%%%%%%%%%%%%%%%%%%%%%%%%%%%%%%%%%%%%
%%%%%%%%%%%%%%%%%%%%%%%%%%%%%%%%%%%%%%%%%%%%%%%%%%%%%%%%%%%%%%%%%%%%%%

\begin{figure}[b]
\centerline{\psfig{figure=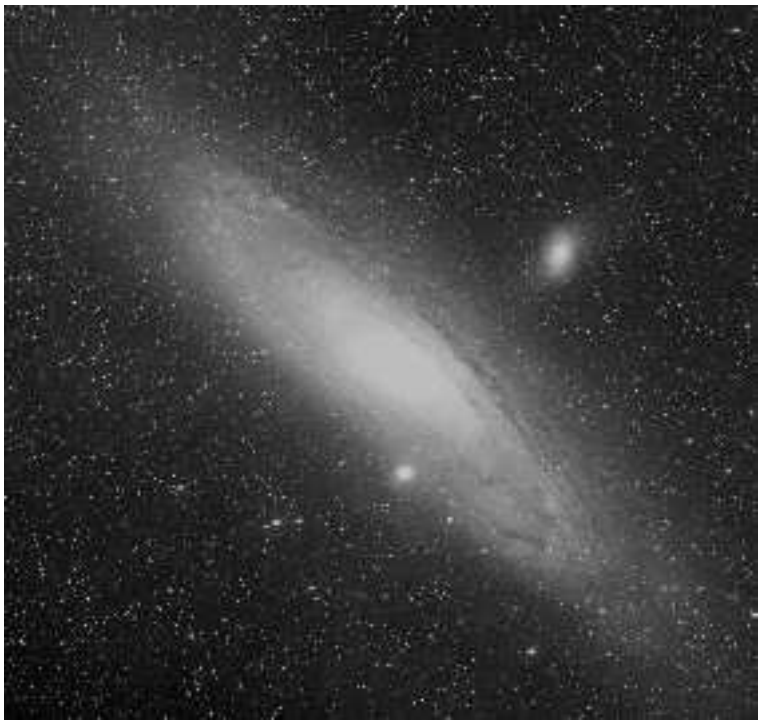,width=10.0 cm}}
\caption{M31, the Andromeda galaxy, the closest spiral galaxy to the
Milky Way at a distance of about $750\,\rm kpc$.
\label{fig:M31}}
\end{figure}

\section{DYNAMICAL EVIDENCE}

\subsection{Rotation Curves of Spiral Galaxies}

\begin{figure}[ht]
\centerline{\psfig{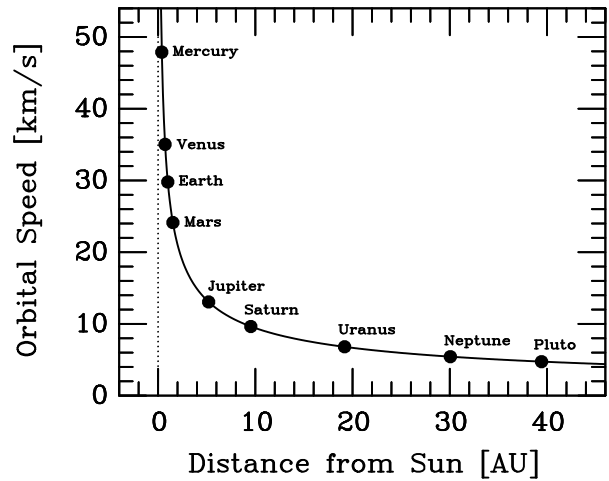}}
\caption{Rotation curve of the solar system which falls off as
$1/\protect\sqrt{r}$ in accordance with Kepler's law.
The astronomical unit (AU) is the Earth-Sun distance of
$1.50\times10^{13}\,\rm cm$.
\label{fig:kepler}}
\end{figure}

Why are astronomers so sure that there are large amounts of dark
matter lurking everywhere in the universe?  The flat rotation curves
of spiral galaxies provide perhaps the most direct and surely the most
impressive evidence. These systems consist of a central bulge and a
very thin disk which is stabilized against collapse by angular
momentum conservation. It is then natural to use the Doppler shift of
spectral lines to obtain a rotation curve, i.e.\ the orbital velocity
of the disk as a function of radius. For the Andromeda galaxy
(Fig.~\ref{fig:M31}), our next-door neighbor at a distance of
about\footnote{Astronomical distances are usually measured in parsec
(pc) where $1\,{\rm pc}=3.26\,\hbox{\rm light-years}
=3.08\times10^{18}\,{\rm cm}$. As a matter of general orientation
note that $1\,\rm pc$ is a typical distance between stars within the
galactic disk, $10\,\rm kpc$ is a typical scale for a galactic disk
(the Sun is at $8.5\,\rm kpc$ from the center of the Milky Way)
galaxies are typically $1\,\rm Mpc$ away from each other, and the
visible universe has a radius of about $3\,\rm Gpc$.} $750\,\rm
kpc$, the rotation curve was first measured by Babcock in
1938~\cite{Babcock}.  Later when it became possible to measure
galactic rotation curves far out into the disk a most unusual behavior
emerged.  The orbital velocity rose roughly linearly from the center
outward until it reached a typical value of around $200\,\rm
km\,s^{-1}$. The rotation curve then stayed flat at this velocity out
to the largest measured radii, a systematic trend clearly
diagnosed as such by Freeman in 1970~\cite{Freeman}. This behavior is
completely unexpected because the surface luminosity of the disk falls
off exponentially with radius~\cite{Freeman}
\begin{equation}\label{eq:expdisk}
I(r)=I_0\,e^{-r/r_{\rm D}},
\end{equation}
where $r_{\rm D}$ is the ``disk scale-length.''  Therefore, one would
expect that most of the galactic mass is concentrated within a few
scale-lengths and that the orbital velocity $v_{\rm rot}$ of the disk
material is determined by this mass just as the orbital velocity of
the planets in the solar system is dominated by the mass of the Sun.
Because in such a system we have $v_{\rm rot}=\sqrt{G_{\rm N} M/r}$
(central mass $M$, Newton's constant $G_{\rm N}$) one expects the
Keplerian $v_{\rm rot}\propto r^{-1/2}$ behavior in analogy to the
solar system~(Fig.~\ref{fig:kepler}).

\begin{figure}[b]
\centerline{\psfig{figure=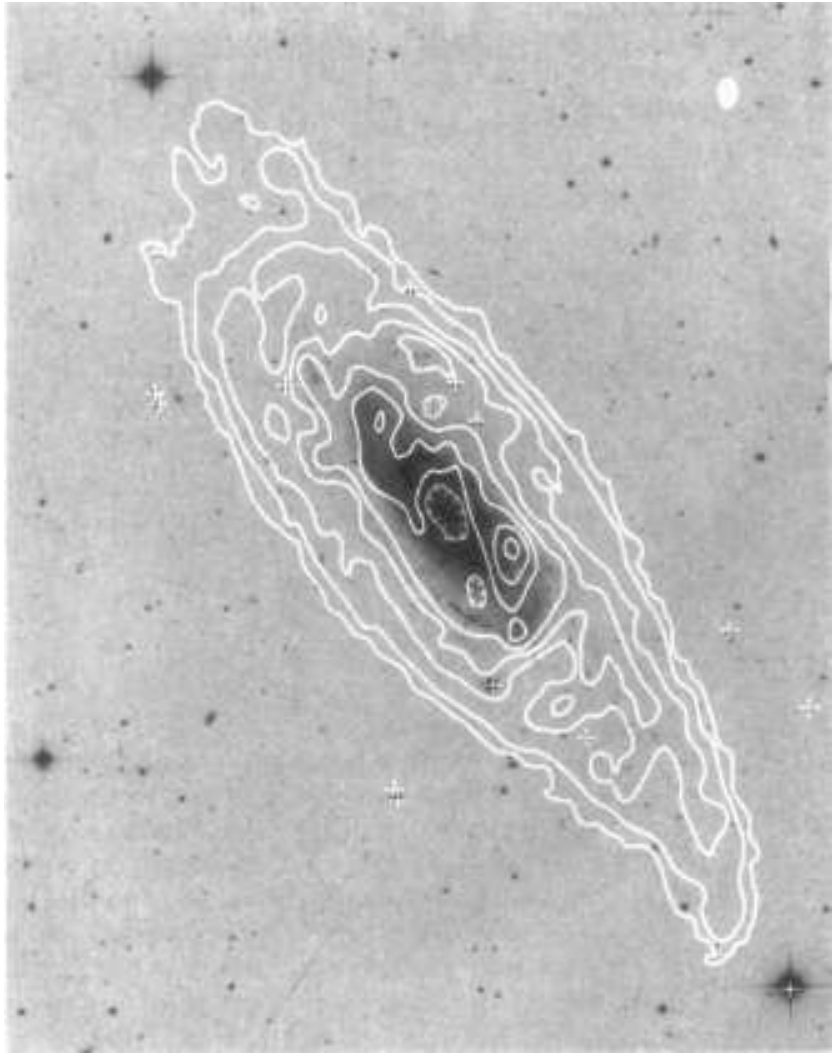,width=11.0 cm}}
\caption{Image of the spiral galaxy NGC~3198 with a 
superimposed contour map of the column density of hydrogen 
gas~\protect\cite{Albada}.
\label{fig:radio}}
\end{figure}

\begin{figure}
\centerline{\psfig{figure=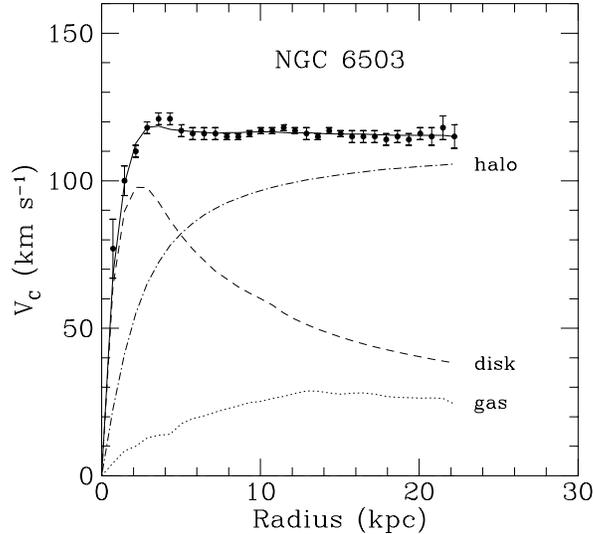,width=8.0 cm}}
\caption{Rotation curve of the spiral galaxy NGC~6503 as established
from radio observations of hydrogen gas in the
disk~\protect\cite{Begeman}.  The last measured point is at $12.8$
disk scale-lengths.  The dashed line shows the rotation curve expected
from the disk material alone, the dot-dashed line from the dark matter
halo alone.
\label{fig:rotation}}
\end{figure}

The non-Keplerian, essentially flat nature of the rotation curves is
supported by systematic optical studies of many spiral
galaxies~\cite{Rubin80,Rubin83}.  The most convincing evidence for
this unexpected behavior, however, arises from radio observations.
Spiral galaxies typically have neutral hydrogen in their disks which
can be observed by its $21\,\rm cm$ line emission. The hydrogen can be
observed to much larger galactic radii than optical tracers
(Fig.~\ref{fig:radio}) so that one can obtain far more extended
rotation curves~\cite{Bosma,Albada,Begeman} than by purely optical
observations which typically stop at $1.5{-}3.5$ disk scale-lengths.
A case in point is the galaxy NGC~6503 where $r_{\rm D}=1.73\,\rm kpc$
while the last measured hydrogen point is at $r=22.22\,{\rm kpc}
=12.8\,r_{\rm D}$. The measured rotation curve is shown in
Fig.~\ref{fig:rotation} together with the relative components ascribed
to the gravity of the disk alone and gas alone.

\newpage

The difference to the rotation curve which is expected from the
luminous material is ascribed to the gravitational effect of dark
matter. A number of strong arguments suggest that this material cannot
be part of the galactic disk itself. First, the distribution of stars
vertically to the galactic disk in our galaxy together with their
vertical velocity dispersion reveals that there cannot be any
significant amount of dark matter confined to the disk, although it
has been the subject of some debate since 1932 if there is {\em
some\/} disk dark matter~\cite{Oort}. Second, a thin self-gravitating
disk is dynamically unstable. Third, the hydrogen of the disk tends to
be vertically far more extended than would be expected if all of the
gravitating matter were in the disk, especially at large
galactocentric radii (``hydrogen flaring''). Fourth, there exist
``polar ring galaxies'' with material orbiting perpendicularly to the
normal disk which appears to trace out a more or less spherical
gravitational potential. (For a review of such arguments
see~\cite{Trimble}.)  An overall picture of spiral galaxies emerges
where the bulge and disk are dynamically subdominant components
immersed in a huge spherical ``halo'' or ``corona'' of dark matter. It
is not crucial that this halo be strictly spherical; the overall
picture does not change if the halo exhibits a significant degree of
oblateness or even triaxiality.

\begin{figure}[b]
\centerline{\psfig{figure=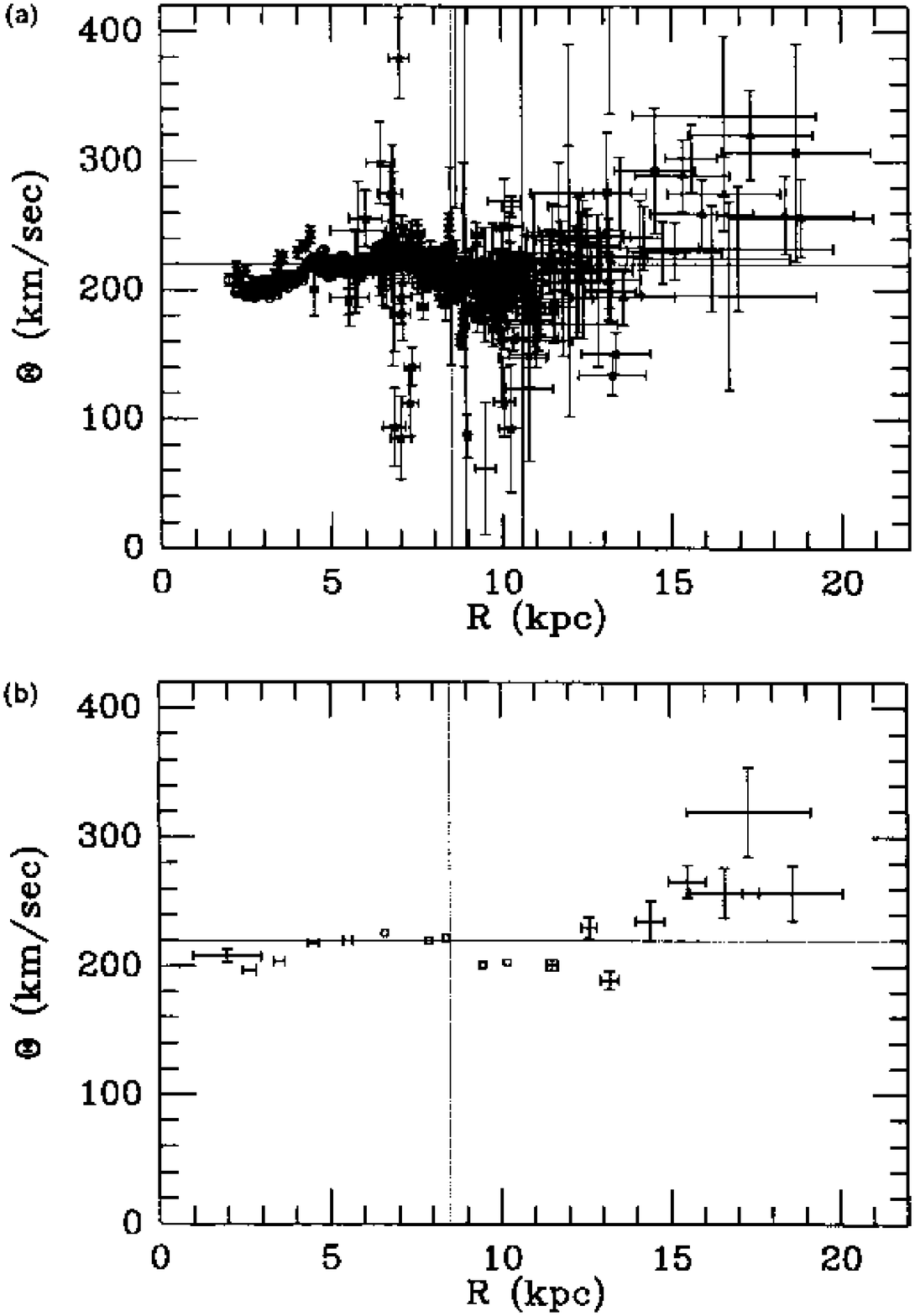,width=8.0 cm}}
\caption{Rotation curve of the Milky Way where $\Theta$ stands for
$v_{\rm rot}$~\protect\cite{Fich}.  The dotted lines represent the
1985 International Astronomical Union values of $v_{\rm rot}=220\,{\rm
km\,s^{-1}}$ at the location of the solar system which is taken to be
at a galactocentric radius of $8.5\,\rm kpc$. The upper panel
represents all data as quoted in Ref.~\protect\cite{Fich}, the lower
panel their smoothed data set.
\label{fig:milky}}
\end{figure}

\newpage

The study of more than a thousand galactic rotation curves reveals
that empirically they can be represented extremely well 
by a ``universal rotation curve'' (URC)~\cite{Salucci}
\begin{eqnarray}\label{eq:URC}
v_{\rm URC}(r)&=&v(r_{\rm opt})
\left\{\left(0.72+0.44\,\log_{10}\frac{L}{L_*}\right)\,
\frac{1.97\,x^{1.22}}{(x^2+0.78^2)^{1.43}}\right.\nonumber\\
&&\hskip3em
+\,\left.\left(0.28-0.44\,\log_{10}\frac{L}{L_*}\right)
\left[1+2.25\left(\frac{L}{L_*}\right)^{0.4}\right]\,
\frac{x^2}{x^2+2.25\,(L/L_*)^{0.4}}\right\}^{1/2},
\end{eqnarray}
where $x\equiv r/r_{\rm opt}$, $L$ is the luminosity of the
galaxy, and the reference luminosity is $L_*\equiv 2.5\times10^{10}
L_\odot$ in the optical $B$-band (blue filter) with $L_\odot$ the
solar luminosity.  The optical radius $r_{\rm opt}$ is defined to
encompass 83\% of the integrated light; for the exponential disk of
Eq.~(\ref{eq:expdisk}) we have $r_{\rm opt}=3.2\,r_{\rm
D}$. Empirically, then, galactic rotation curves depend on only two
parameters, the total luminosity and the optical radius.

Galaxies presumably form by the infall of material in an overdense
part of the universe which has decoupled from the overall cosmic
expansion. The dark matter is supposed to undergo ``violent
relaxation'' and form a ``virialized system.''  This picture has led
to a simple model of dark-matter halos as ``modified isothermal
spheres.'' The radial density profile is taken to be
\begin{equation}\label{eq:isosphere}
\rho(r)=\frac{v_\infty^2}{4\pi G_{\rm N}r_{\rm c}^2}\,
\frac{r_{\rm c}^2}{r_{\rm c}^2+r^2},
\end{equation}
where $r_{\rm c}$ is a core radius and $v_\infty$ the plateau value of
the flat rotation curve. This sort of model is consistent with the
universal rotation curve of Eq.~(\ref{eq:URC}) if one disentangles the
luminous-matter contribution from the total rotation curve. At large
radii such a distribution leads to a strictly flat rotation curve.

The URC reveals that the more luminous galaxies are dominated
by luminous matter to relatively large radii while the fainter ones
are more dominated by dark matter. The faintest (smallest) galaxies
are dominated by dark matter even in their central regions. Therefore,
these systems are better laboratories than bright spirals to test
theories of galaxy formation. Actually, the best measured rotation
curve is that of the dwarf spiral DDO~154 which extends out to about
20 disk scale lengths. In such systems the rotation curve falls off at
large radii; their dark matter density profile
is well represented by~\cite{Burkert}
\begin{equation}
\rho(r)=\rho_0\,\frac{r_0^3}{(r+r_0)(r^2+r_0^2)},
\end{equation}
where $\rho_0$ is the central density and $r_0$ a core radius. The
integral mass diverges only logarithmically with radius. The large-$r$
behavior of this model is predicted by recent high-resolution $N$-body
simulations of galaxy formation in a cold dark matter
cosmology~\cite{Navarro}.  Towards the galactic center, however, these
simulations predict a density cusp of the form
$[r\,(r^2+r_0^2)]^{-1}$, in apparent contradiction with the
observations. This discrepancy is a possible problem for cold dark
matter cosmologies~\cite{Moore} even though the reality of the
discrepancy has recently been questioned~\cite{Kravtsov}.

For the purpose of the direct detection of dark matter our own Milky
Way is the most interesting system. Its rotation curve is far more
difficult to obtain than that of an external galaxy because we can see
only part of it (most is obscured by dust in the disk) and it is
difficult to obtain reliable galactocentric distances for the tracers.
Still, the rotation curve of Fig.~\ref{fig:milky} shows that the Milky
Way conforms to the usual picture.  The approximate plateau value for
the rotation velocity is $220\,\rm km\,s^{-1}$.  For dark matter
search experiments the most critical quantity is the dark matter
density in the solar neighborhood. The canonical value usually adopted
for the interpretation of the experiments is
\begin{equation}\label{eq:halodensity}
\rho_{\rm DM}=300\,\rm MeV\,cm^{-3}.
\end{equation}
It must be kept in mind, however, that this number depends on the
model adopted for the galactic dark-matter halo and thus is uncertain
to within, perhaps, a factor of two~\cite{JKG96}.

\subsection{Cosmic Density Contribution of Galaxies}

Another important question is how much the total masses of galaxies
contribute to the overall density of the universe. It is usually
expressed in terms of the cosmic critical density~\cite{KolbTurner}
\begin{equation}
\rho_{\rm crit}\equiv\frac{3 H^2_0}{8\pi G_{\rm N}}
=h^2\,1.88\times10^{-29}\,\rm g\,cm^{-3},
\end{equation}
where $H_0$ is the present-day Hubble expansion parameter.
It is usually written as
\begin{equation}\label{eq:hdef}
H_0=h\,100\,\rm km\,s^{-1}\,Mpc^{-1}
\end{equation}
in terms of the dimensionless parameter $h$ which appears in various
powers in most quantities of cosmological interest. Observationally it
lies in the range $0.4\alt h\alt 1.0$ with
\begin{equation}
0.5\alt h\alt 0.8
\end{equation}
the currently most favored interval~\cite{Hogan}. The average
contribution $\rho$ of various matter components to the cosmic density
is usually expressed by the parameter
\begin{equation}
\Omega\equiv\rho/\rho_{\rm crit}.
\end{equation}
In the framework of the usual Friedmann-Lema\^itre-Robertson-Walker
cosmology~\cite{KolbTurner} the spatial cosmic geometry is Euclidean
for $\Omega=1$ (``flat universe''), the spatial curvature is negative
for $\Omega<1$ (``open universe''), and it is positive for $\Omega>1$
(``closed universe'').

The contribution of galaxies to $\Omega$ is related to the luminosity
density of the universe which is found to be
$(1.7\pm0.6)\times10^8\,h\,L_\odot{\rm Mpc}^{-3}$ in the $V$ (visual)
spectral band~\cite{LumDens}.  This luminosity density can be
translated into a mass density by a multiplication with the
mass-to-light ratio $M/L$ of a given class of systems, often denoted
by $\Upsilon$ (upsilon).  Mass-to-light ratios are usually expressed
in solar units $M_\odot/L_\odot$ so that for the Sun $\Upsilon=1$.
Therefore, the cosmic mass density is
$\Omega=(6.1\pm2.2)\times10^{-4}\, h^{-1}\,\Upsilon_V$.  The
luminosity of stars depends sensitively upon their mass and their
stage of evolution.  Stellar populations for which the mass and
luminosity can be determined independently include globular clusters
and the disks of spiral galaxies which have an $\Upsilon$ of a few.
The stars in the solar neighborhood have $\Upsilon\approx5$. Taking
this as a representative value we find for the luminous mass density
of the universe $\Omega_{\rm lum}h\approx0.003$. Several methods give
values which are consistent with the range~\cite{CopiSchramm}
\begin{equation}\label{eq:lumdens}
0.002\alt\Omega_{\rm lum}h\alt0.006.
\end{equation}
Therefore, the luminous matter alone is far from the cosmic critical
density.  

The mass-to-light ratios of galactic haloes are typically at least
around $30\,h$ as far as the measured rotation curves reach, giving a
cosmic mass density of at least
\begin{equation}
\Omega_{\rm gal}\agt0.03{-}0.05.
\end{equation}
The flat rotation curves indicate that their integral mass increases
as $M(r)\propto r$. Because the rotation curves tend to stay flat out
to the largest radii where tracers are available, the true size of
galactic dark matter halos and thus the total cosmic mass in galaxies
is not well known. Estimating the extent of dark-matter haloes from
satellite dynamics yields $\Omega_{\rm gal}h=0.2{-}0.5$
\cite{satellites}.

%%%%%%%%%%%%%%%%%%%%%%%%%%%%%%%%%%%%%%%%%%%%%%%%%%%%%%%%%%%%%%%%%%%%%%
%%%%%%%%%%%%%%%%%%%%%%%%%%%%%%%%%%%%%%%%%%%%%%%%%%%%%%%%%%%%%%%%%%%%%%

\begin{figure}[b]
\centerline{\psfig{figure=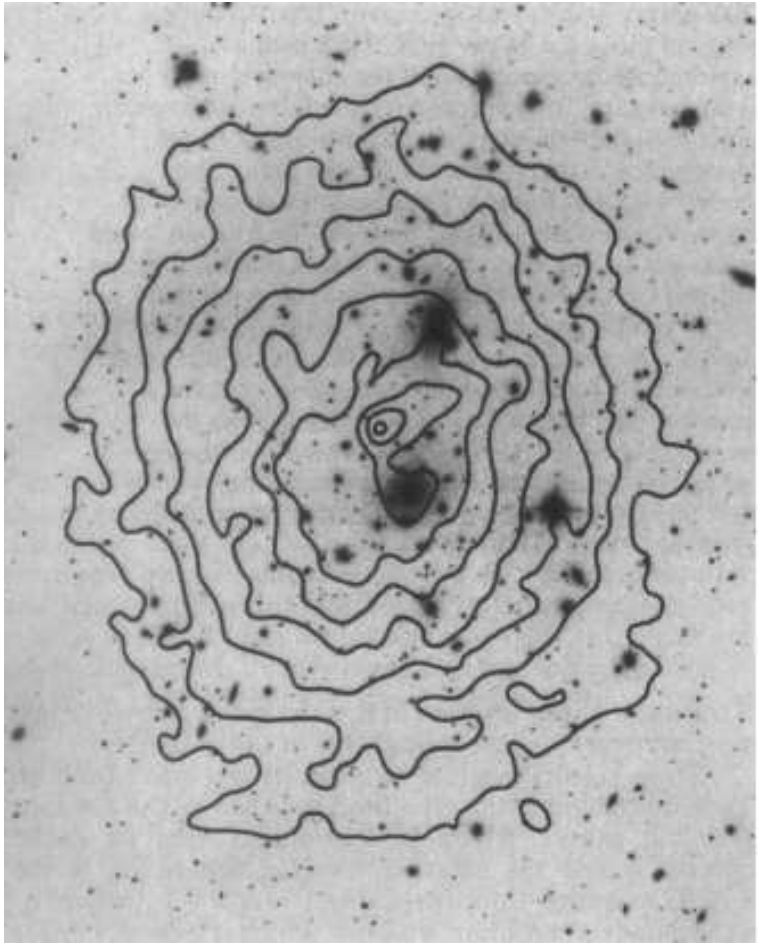,width=10.0 cm}}
\caption{Coma cluster of galaxies.  Contour map of the x-ray surface
brightness measured by the Einstein satellite superimposed on an
optical image. (Picture by William Forman and Christine Jones,
Harvard-Smithonian Center for Astrophysics, here reproduced from
Ref.~\protect\cite{Tremaine}.)
\label{fig:xray}}
\end{figure}

\subsection{Clusters of Galaxies}
\label{sec:ClustersOfGalaxies}

Clusters of galaxies are the largest gravitationally bound systems in
the universe. We know today several thousand clusters; they have
typical radii of $1.5\,\rm Mpc$ and typical masses of
$5\times10^{14}\,M_\odot$.  Zwicky first noted in 1933 that these
systems appear to contain large amounts of dark matter~\cite{Zwicky}.
He used the virial theorem which tells us that in a gravitationally
bound system in equilibrium
\begin{equation}
2\langle E_{\rm kin}\rangle=-\langle E_{\rm grav}\rangle
\end{equation}
where $\langle E_{\rm kin}\rangle=\frac{1}{2}m\langle v^2\rangle$ is
the average kinetic energy of one of the bound objects of mass $m$ and
$\langle E_{\rm grav}\rangle= - m G_{\rm N}\langle M/r \rangle$ is the
average gravitational potential energy caused by the other bodies.
Measuring $\langle v^2\rangle$ from the Doppler shifts of the spectral
lines and estimating the geometrical extent of the system gives one
directly an estimate of its total mass $M$.  As Zwicky stressed, this
``virial mass'' of the clusters far exceeds their luminous matter
content, typically leading to a mass-to-light ratio of around 300.
From current estimates for virial cluster masses one finds for the
cosmic matter density~\cite{Carlberg}
\begin{equation}
\Omega_{\rm M}=0.24\pm0.05\pm0.09,
\end{equation}
where the first uncertainty is a statistical $1\sigma$ error while the
second is an estimate of systematic uncertainties. It was assumed that
the average cluster $M/L$ is representative for the entire universe
which is not to be taken for granted as most galaxies are actually not
in clusters but in the general field.

\begin{figure}[b]
\centerline{\psfig{figure=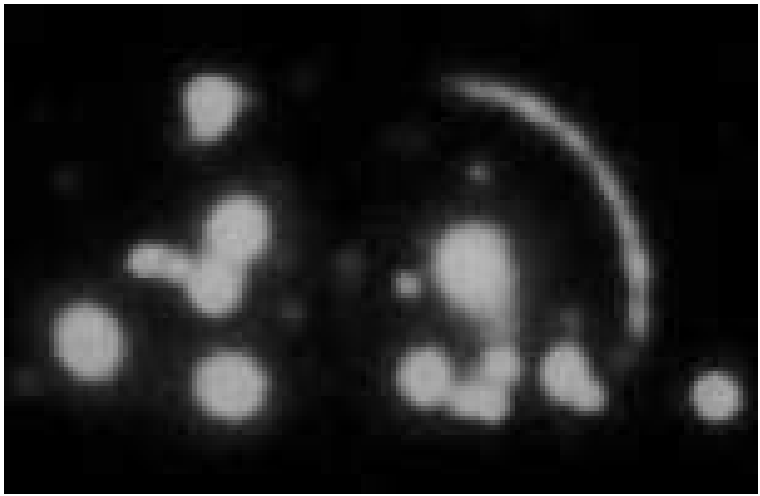,width=10.0 cm}}
\caption{Giant arc in the cluster Cl 2244-02 which is at a redshift of
$z=0.33$ while the source which is imaged as an arc is at $z=2.24$.
\label{fig:arc}}
\end{figure}

After the mid 1960s when x-ray telescopes became available it turned
out that galaxy clusters are the most powerful x-ray sources in the
sky~\cite{Sarazin}.  The emission is extended over the entire cluster
(Fig.~\ref{fig:xray}) and thus reveals the presence of large amounts
of ``x-ray gas,'' a hot plasma ($T=10^7{-}10^8\,\rm K$) where x-rays
are produced by electron bremsstrahlung. Assuming this gas to be in
hydrostatic equilibrium one may again apply essentially the virial
theorem (with the gas particles being the test bodies) to estimate the
total cluster mass, generally giving approximate agreement (within a
factor of 2) with the virial mass estimates. The total mass in the
x-ray gas is typically in the $10{-}20\%$ range~\cite{WhiteFabian},
i.e.\ clusters contain more baryonic matter in the form of hot gas
than in the form of stars in galaxies. This large baryon fraction
relative to the total cluster mass, if taken to be representative of
the entire universe, indicates that the amount of nonbaryonic dark
matter exceeds the cosmic baryon content only by a factor of around
$10$, a finding with important cosmological
ramifications~\cite{White93} as we shall see below.

In the mid 1980s one began to observe huge arc-like features in galaxy
clusters~\cite{arcs,Fort94} with one prominent example shown in
Fig.~\ref{fig:arc}. The cluster galaxies and these ``giant arcs'' are
at very different cosmological redshifts and thus at very different
distances. The standard interpretation is that the arc is the image of
a distant background galaxy which is almost lined up with the cluster
so that it appears distorted and magnified by the gravitational lens
effect~\cite{Paczynski87}. A source and a gravitational deflector
which are precisely lined up would give rise to a ring-like image
(``Einstein ring''); the giant arcs are essentially partial Einstein
rings. The cluster mass estimates derived from this interpretation,
again, reveal large amounts of dark matter in rough agreement
(approximate factor of~2) with the virial mass estimates, even though
the lensing masses tend to be systematically larger~\cite{NB95}.

\begin{figure}[t]
\centerline{\psfig{figure=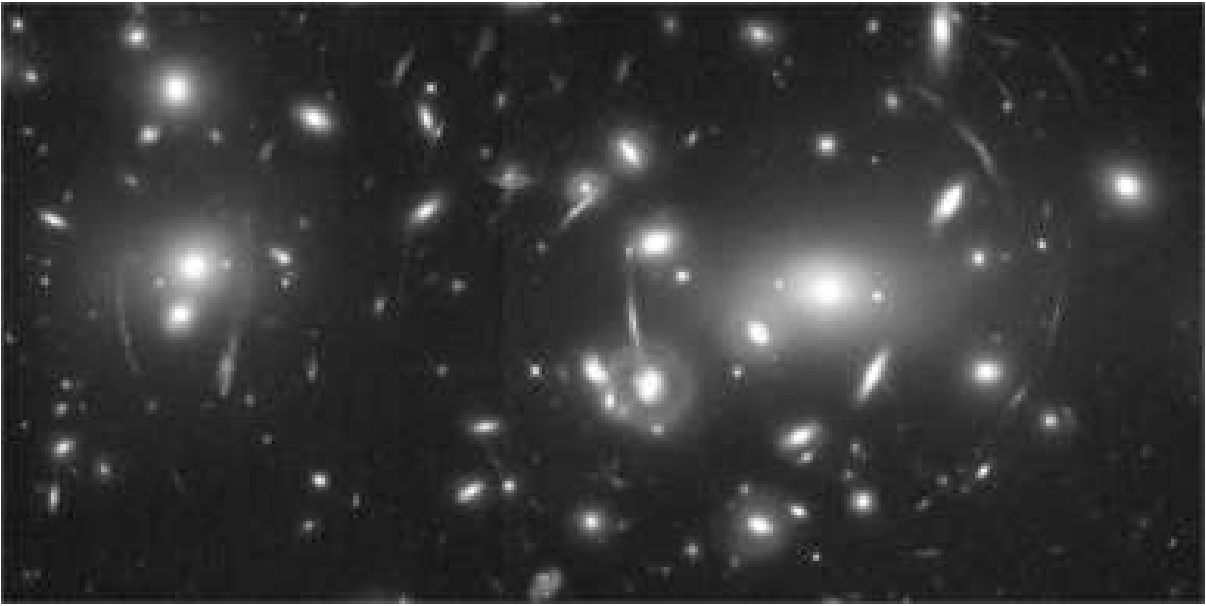,width=\hsize}}
\caption{Hubble Space Telescope (HST) image of the cluster Abell 2218,
showing a number of arcs and arclets around the two centers of the 
cluster (NASA HST Archive).
\label{fig:arclet}}
\end{figure}

\begin{figure}[ht]
\centerline{\psfig{figure=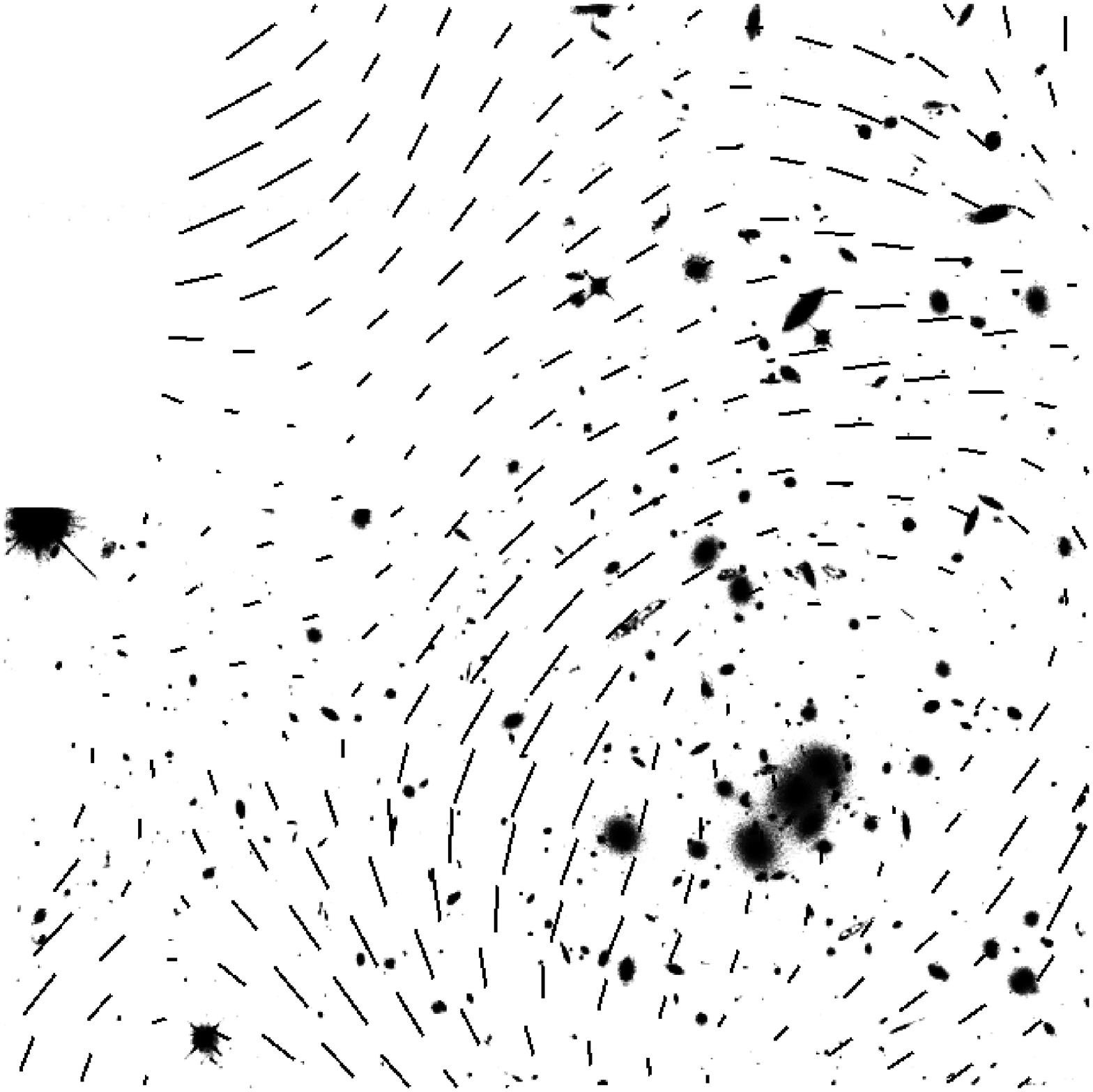,width=6cm}
\hskip0.5cm\psfig{figure=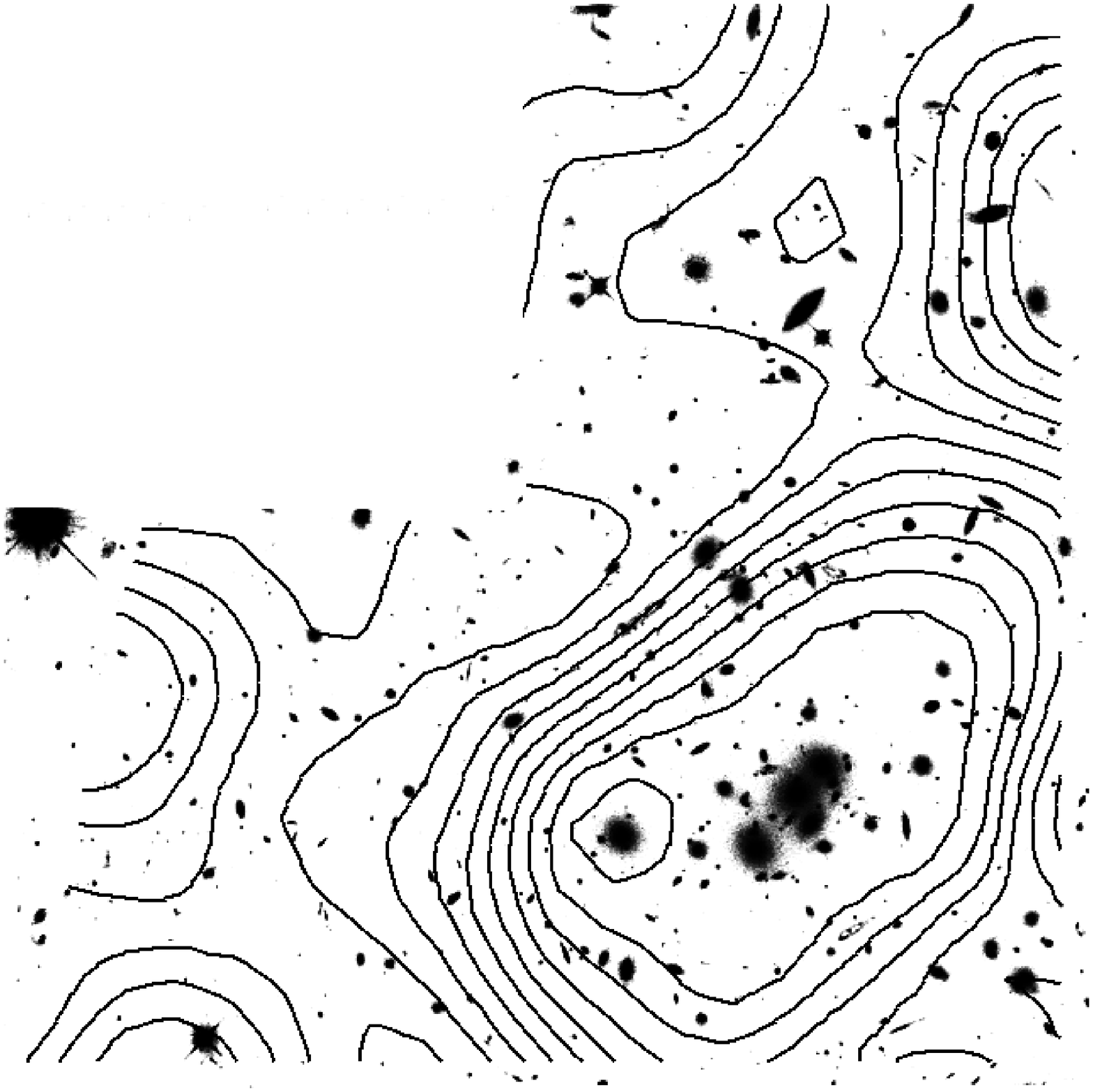,width=6cm}}
\caption{HST image of the cluster Cl~0024, overlaid on the left with
the shear field obtained from an observation of arclets with the
Canada-France Hawaii Telescope (Y.~Mellier and B.~Fort),
and on the right with the reconstructed surface-mass density
determined from the shear field (C.~Seitz et~al.). (Figure from
Ref.~\protect\cite{NB95}.)
\label{fig:shear}}
\end{figure}

While the appearance of giant arcs requires a special alignment
between source and lens, the image of every background galaxy in the
vicinity of a given cluster will be distorted, causing the appearance
of innumerable ``arclets'' (Fig.~\ref{fig:arclet}).  This ``weak
lensing effect'' allows for a systematic study of cluster mass
distributions~\cite{Tyson}. One uses the statistical distributions of
arclets to reconstruct the shear field of gravitational image
distortions and from there one can derive cluster mass distributions
(Fig.~\ref{fig:shear}). This approach to mapping out cluster dark
matter has turned into a new topical field of astronomical research in
its own right~\cite{Fort94,NB95,Tyson,Geiger97}.

\begin{figure}[b]
\centerline{\psfig{figure=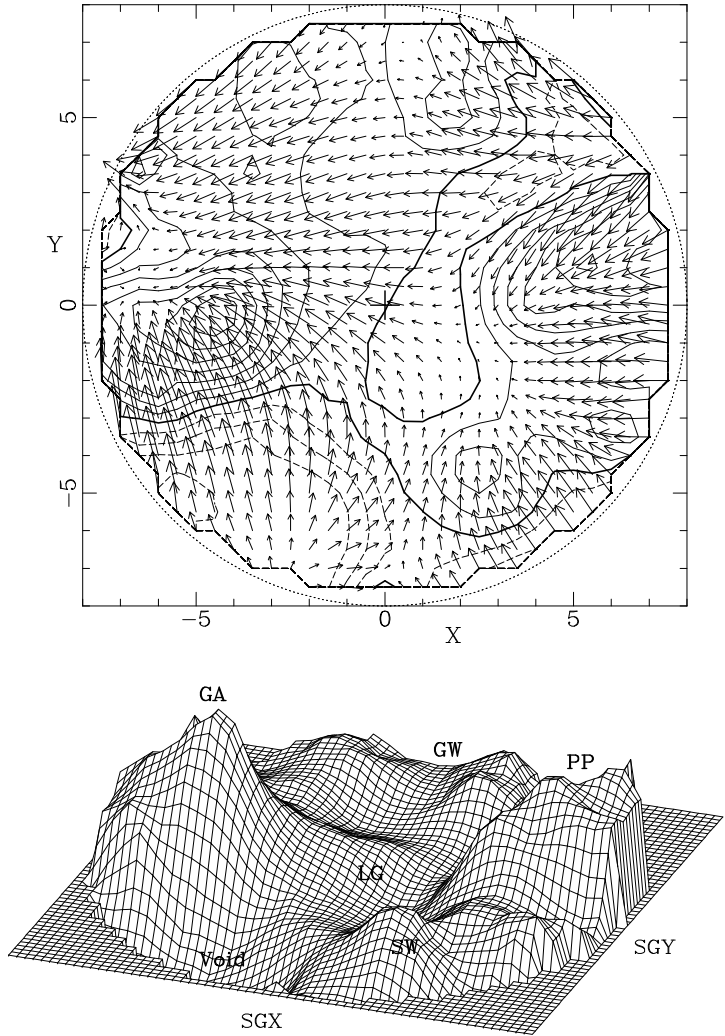,width=8.5 cm}}
\vskip0.5cm
\centerline{\psfig{figure=fig11b.eps,width=9.5cm,angle=270}}
\caption{The fluctuation fields of velocity and mass-density in the
  Supergalactic Plane as recovered by the POTENT method from the
  Mark~III velocities~\protect\cite{MarkIII} of about 3000 galaxies
  with $12\,h^{-1}\,\rm Mpc$ smoothing~\protect\cite{Dekel}. The
  vectors are projections of the 3-D velocity field in the frame of
  the cosmic microwave background. Dashed contours mark
  underdensities, solid ones overdensities. Distances are in the space
  of Hubble recession velocities in units of $1000\,\rm km\,s^{-1}$
  which corresponds to $h^{-1}\,10\,\rm Mpc$, i.e.\ the reconstruction
  goes out to a distance of $h^{-1}\,80\,\rm Mpc$.  The marked
  structures are the local group (LG), the ``great attractor'' (GA),
  the Coma ``great wall'' (GW), the Perseus-Pisces (PP) region, and
  the ``southern wall'' (SW).
\label{fig:dekel}}
\end{figure}

\subsection{Large-Scale Motion}

On scales larger than clusters the motion of galaxies is dominated by
the overall cosmic expansion. Still, they exhibit ``peculiar
velocities'' relative to the overall cosmic flow.  For example, our
own group of galaxies moves with a speed of $627\pm22\,\rm km\,s^{-1}$
relative to the reference frame defined by the cosmic microwave
background radiation. For external galaxies the determination of
peculiar velocities requires the determination of their redshifts and
an independent measure of distance. A homogeneous catalog of about
3000 galaxies (Mark III catalog~\cite{MarkIII}) has recently been
completed for this purpose.

In the context of the standard gravitational instability theory of
structure formation the peculiar motions are attributed to the action
of gravity over the age of the universe, caused by the matter density
inhomogeneities which give rise to the formation of structure. The
observed large-scale velocity fields together with the observed galaxy
distributions can then be translated into a measure for the
mass-to-light ratio which is necessary to explain the large-scale
flows.  An example for the reconstruction of the matter density field
in our cosmological neighborhood from the observed velocity field by
means of the POTENT method is shown in Fig.~\ref{fig:dekel}. The
cosmic matter density inferred by such methods is~\cite{Dekel}
\begin{equation}\label{eq:flowmass}
\Omega_{\rm M}>0.3,
\end{equation}
which is claimed to be a 95\% C.L.\ lower bound. Related methods which
are more model-dependent give even larger estimates.

\subsection{Cosmic Age}

The dynamical density measured on galactic scales up to those of
large-scale flows provide a lower limit to the average cosmic matter
density $\Omega_{\rm M}$. Naturally, the true $\Omega_{\rm M}$ could
be larger than indicated by Eq.~(\ref{eq:flowmass}). Less reliable
arguments concerning the dynamics of the large-scale structure and
large-scale flows already point to values for $\Omega_{\rm M}$ not
much below the critical value~1 \cite{Dekel}.  The cosmic matter
density determines the dynamical evolution of the universe through the
Friedmann equation. Therefore, a critical measure of $\Omega_{\rm M}$
is provided by the cosmic age $t_0$ as infered from the ages of the
oldest stars in conjunction with measurements of the present-day
expansion rate $H_0$.

\begin{figure}[b]
\centerline{\psfig{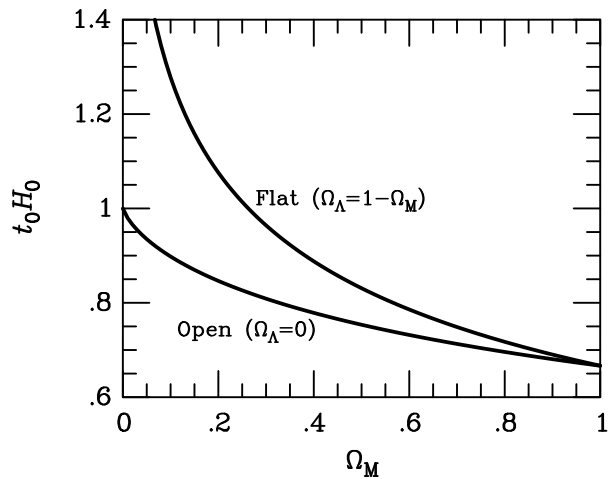}}
\caption{Age of the universe as a function of matter content
according to Eq.~(\protect\ref{eq:age1}) for an open, matter dominated
universe and according to Eq.~(\protect\ref{eq:age2}) for a flat
universe with $\Omega_\Lambda=1-\Omega_{\rm M}$. 
\label{fig:age}}
\end{figure}

The known or conjectured forms of radiation (cosmic microwave photons
and other electromagnetic background radiations, massless neutrinos,
gravitational waves) are thought to contribute only $\Omega_{\rm
rad}h^2\approx3\times10^{-5}$ 
which shall be ignored in the present discussion.
If we thus assume for the moment that the total cosmic energy density
$\Omega_{\rm tot}$ is essentially identical with the matter density
$\Omega_{\rm M}$, and if we assume in addition that $\Omega_{\rm
M}\leq 1$, the relationship between age and matter content
is~\cite{KolbTurner} (Fig.~\ref{fig:age})
\begin{equation}\label{eq:age1}
H_0 t_0=\frac{\Omega_{\rm M}}{2(1-\Omega_{\rm M})^{3/2}}\,
\left[\frac{2}{\Omega_{\rm M}}\,(1-\Omega_{\rm M})^{1/2}
-{\rm Acosh}\left(\frac{2}{\Omega_{\rm M}}-1\right)\right]\,.
\end{equation}
 For $\Omega_{\rm M}=1$ one finds the
well-known limit $H_0 t_0=2/3$.

For a long time there was an ``age crisis'' for the universe in that
the oldest stars seemed older than its expansion age. However, recent
modifications of the stellar input physics (equation of state,
opacities, etc.) and particularly the new Hipparcos calibration of
stellar distances have led to a revision of the age estimates for the
oldest globular clusters to $10{-}13\,\rm Gyr$ \cite{StellarAges}.
Moreover, estimates for the Hubble constant have come down to about
$0.5\alt h\alt0.8$ \cite{Hogan}, leading to an allowed range of
$0.5\alt H_0 t_0\alt 0.8$ which includes the critical-universe value
$2/3$ without any problems.

The critical value $\Omega_{\rm tot}=1$ for the total cosmic mass and
energy density, corresponding to an overall Euclidean (flat) spatial
geometry, is strongly favored to avoid a fine-tuning problem of cosmic
inital conditions. In an expanding universe $\Omega_{\rm tot}$ evolves
quickly away from 1 towards either $0$ or $\infty$ so that the
near-flatness of the present-day universe suggests $\Omega_{\rm
tot}=1$ as an exact identity. Moreover, inflationary models of the
early universe generically produce a flat geometry even though one may
construct fine-tuned models which can circumvent this as an
absolute prediction.

However, even if the universe is flat one is not assured that
$\Omega_{\rm tot}$ is dominated by matter; a cosmological constant
$\Lambda$ is also conceivable.  This hypothesis periodically comes and
goes in cosmology.  In modern terms $\Lambda$ arises as the vacuum
energy of quantum fields and as such poses the opposite problem, i.e.\ 
why is it not much larger than the cosmologically allowed value.  Its
observed smallness remains unexplained and no compelling reason is
known why $\Lambda$ should be exactly zero~\cite{Weinberg}. Therefore,
from the cosmological perspective $\Lambda$ and its contribution
$\Omega_\Lambda$ to the total energy density remains a free parameter.

A cosmological constant (vacuum energy) differs from matter in a
number of important ways. $\Omega_\Lambda$ can be both positive or
negative while $\Omega_{\rm M}$ is always positive.  Vacuum energy is
homogeneously distributed and thus cannot be measured dynamically on
scales of galaxies, clusters, and so forth; it only affects the global
dynamics of the universe. The most counter-intuitive property is that
vacuum energy is not diluted by the cosmic expansion. The normal
matter density $\rho_{\rm M}$ is conserved in a co-moving volume and
thus is diluted as $R^{-3}$ with the time-dependent cosmic
scale-factor $R$ while the vacuum energy density $\rho_{\rm
vac}$ is constant so that its contribution grows as $R^3$ in a
co-moving volume. Therefore, if there is any vacuum energy it
dominates the dynamics at late times.

Pragmatically, then, the choice of cosmological models is between a
matter-dominated open universe with no cosmological constant, a
matter-dominated flat universe, and a flat universe with a certain
cosmological-constant contribution ($\Omega_{\rm tot}=\Omega_{\rm
M}+\Omega_\Lambda=1$). In this latter case the age 
is~\cite{KolbTurner}
\begin{equation}\label{eq:age2}
H_0 t_0=\frac{2}{3}\,\frac{1}{(1-\Omega_{\rm M})^{1/2}}
\ln\left(\frac{1+(1-\Omega_{\rm M})^{1/2}}{\Omega_{\rm M}^{1/2}}
\right),
\end{equation}
also shown in Fig.~\ref{fig:age}. For the same $\Omega_{\rm M}$ this
gives a larger expansion age than an open matter-dominated universe,
and a much larger expansion age than a flat matter-dominated universe.
Until recently the age crisis together with a number of arguments
related to structure formation seemed to point toward a cosmological
constant~\cite{CosmologicalConstant}, but today this case is far less
compelling even if it can still be argued~\cite{Krauss97}.

%%%%%%%%%%%%%%%%%%%%%%%%%%%%%%%%%%%%%%%%%%%%%%%%%%%%%%%%%%%%%%%%%%%%%%
%% Section III %%%%%%%%%%%%%%%%%%%%%%%%%%%%%%%%%%%%%%%%%%%%%%%%%%%%%%%
%%%%%%%%%%%%%%%%%%%%%%%%%%%%%%%%%%%%%%%%%%%%%%%%%%%%%%%%%%%%%%%%%%%%%%

\section{ASTROPHYSICAL CONSTRAINTS}

\subsection{Big-Bang Nucleosynthesis}

There are a number of strong astrophysical constraints on the possible
nature of the dark matter that appears to dominate the dynamics of the
universe on galactic scales and above. The first natural question is
if this matter could not be just normal matter in some nonluminous
form, perhaps stellar remnants such as neutron stars or black holes or
molecular hydrogen clouds which are difficult to measure. 

\begin{figure}[b]
\centerline{\psfig{figure=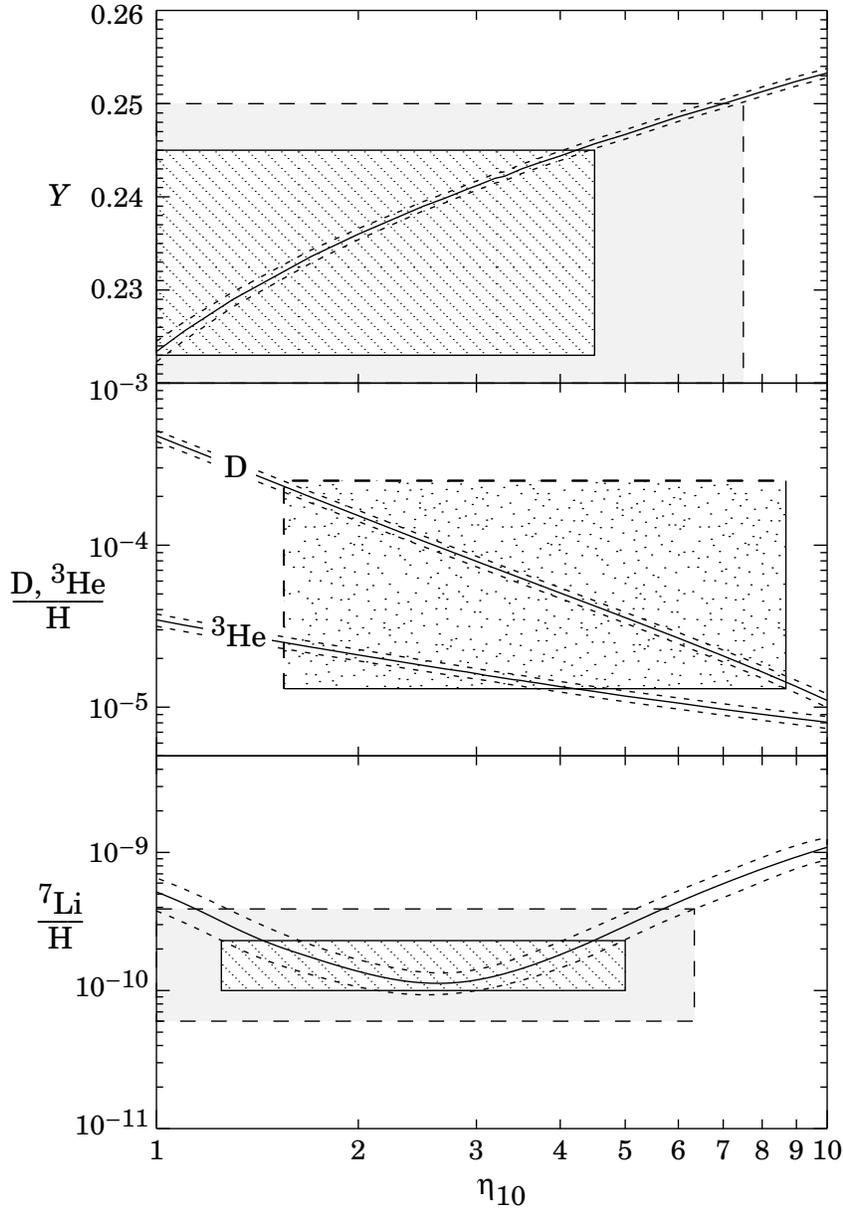,width=11cm}}
\caption{Light-element abundances as a function of the
baryon-to-photon ratio in the standard big-bang
scenario~\protect\cite{OliveSchramm}.  The solid lines show the
standard predictions with their errors due to nuclear cross-section
uncertainties indicated by dashed lines. The boxes indicate the
current observational situation where the big shaded boxes are found
when the systematic uncertainties are pushed to their plausible
limits.
\label{fig:bbn}}
\end{figure}

However, the overall baryonic content of the universe is strongly
constrained by big-bang nucleosynthesis (BBN). When the early universe
cooled below a temperature of about $1\,\rm MeV$ the $\beta$
equilibrium between protons and neutrons froze out, and shortly
afterward all the remaining neutrons together with the ambient protons
formed $^4\rm He$ and traces of deuterium, $^3\rm He$, and $^7\rm Li$
\cite{KolbTurner}. Within the standard big-bang picture the predicted
abundances depend only on one unknown comological parameter, the
baryon number fraction relative to the present-day number density of
cosmic microwave background photons, $\eta\equiv n_B/n_\gamma$.  It is
usually parametrized as $\eta_{10}\equiv \eta/10^{-10}$ and then gives
\begin{equation}
\Omega_{\rm B} h^2= 3.73\times10^{-3}\,\eta_{10}
\end{equation}
for the baryonic mass fraction of the universe. The standard
predictions for the light-element abundances as a function of
$\eta_{10}$ are shown in Fig.~\ref{fig:bbn}.

The main problem is to get an empirical handle at the primordial
light-element abundances from observations in the present-day
universe. The current situation is somewhat confusing in that various
measurements with their claimed uncertainties are not necessarily
mutually consistent. In principle, the most sensitive ``baryon meter''
is the deuterium abundance. Recently, abundance measurements in
intergalactic hydrogen clouds have become possible by observing
deuterium and hydrogen absorption lines from quasars as light
sources. While these measurements hold a great deal of promise toward
a precision determination of the primordial deuterium abundance and
thus of $\eta_{10}$, the current set of results give both
high~\cite{highD} and low~\cite{lowD} values of ${\rm D/H}\approx
2\times10^{-4}$ and $(2.3\pm0.3)\times10^{-5}$, respectively, which
are mutually inconsistent unless the baryon distribution is vastly
inhomogeneous on large scales.

\begin{figure}[ht]
\centerline{\psfig{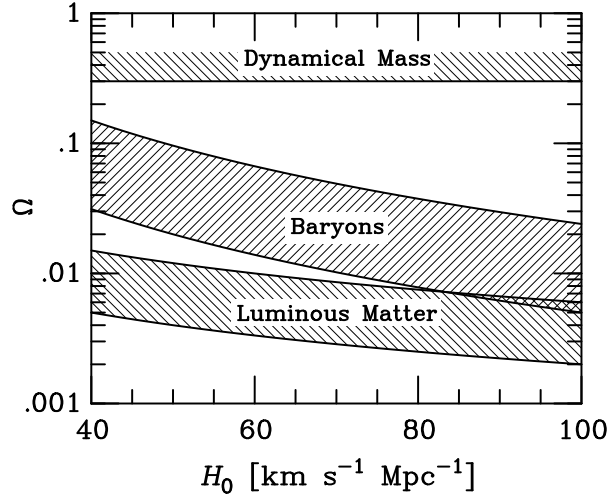}}
\caption{Observed cosmic matter components as a function of the
assumed present-day Hubble expansion parameter. The density range of
luminous matter was given in Eq.~(\protect\ref{eq:lumdens}), the lower
limit to the dynamical mass density in
Eq.~(\protect\ref{eq:flowmass}), and the baryonic density 
inferred from BBN in
Eq.(\protect\ref{eq:bbndens}).
\label{fig:matter}}
\end{figure}

In Fig.~\ref{fig:bbn} the current observational situation is indicated
by error boxes. Olive and Schramm~\cite{OliveSchramm} thus derive a
currently allowed range $1.5<\eta_{10}<6.3$ which implies
\begin{equation}\label{eq:bbndens}
0.005<\Omega_{\rm B}h^2<0.024.
\end{equation}
This allowed range for $\Omega_{\rm B}$ is depicted in
Fig.~\ref{fig:matter} as a function of the Hubble expansion parameter
together with the luminous mass density of Eq.~(\ref{eq:lumdens}) and
the lower dynamical mass limit Eq.~(\ref{eq:flowmass}) from the
analysis of large-scale coherent flows.

The currently favored range for $H_0$ is between $50$ and $80\,\rm
km\,s^{-1}\,Mpc^{-1}$ \cite{Hogan}. Therefore, Fig.~\ref{fig:matter}
reveals that there is a gap between the cosmic baryon density and both
the luminous matter density and the total matter density. Accepting
this result implies that there must be a significant fraction of
``dark baryons'' in the universe which never made their way into
galaxies and stars, and also lots of nonbaryonic dark matter which is
of a completely different physical nature.

\subsection{X-Ray Clusters}
\label{sec:xrayclusters}

We have already discussed in Sec.~\ref{sec:ClustersOfGalaxies} that
galaxy clusters contain a large fraction of baryons in the form of hot
x-ray gas. Estimates of the baryon fraction in clusters relative to
their total mass lead to a baryon fraction of these systems
of~\cite{WhiteFabian,White93}
\begin{equation}
f_{\rm B}h^{3/2} =0.03{-}0.08,
\end{equation}
where higher and lower values are also found in certain cases
\cite{clusterbaryons}. If this cluster baryon fraction is taken to be
representative of the entire universe, and if one uses the
BBN-indicated baryon density of Eq.~(\ref{eq:bbndens}), the cosmic
dark-matter density appears to be less than the critical value 1, but
the current evidence is not strong enough to definitively exclude
$\Omega_{\rm M}=1$ on these grounds.  Still, a low-$\Omega$ universe
is favored by a combination of the BBN baryon content of the universe
and the high cluster baryon fractions, a finding sometimes
referred to as the cluster ``baryon crisis.''

\subsection{Structure Formation}
\label{sec:structureformation}

\begin{figure}[b]
\centerline{\psfig{figure=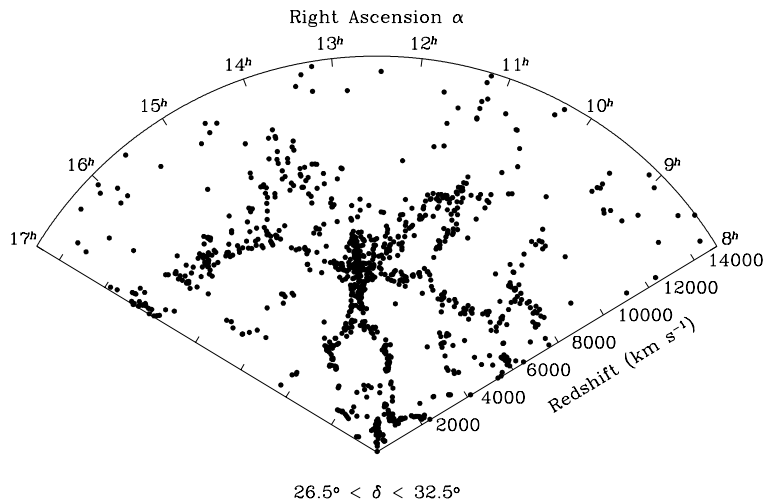,width=10cm}}
\caption{``A slice of the universe:'' The galaxy distribution from the
CfA redshift survey~\protect\cite{Lapparent}. A redshift (apparent
recession velocity) of $100\,\rm km\,s^{-1}$ corresponds to a distance
of $h^{-1}\,\rm Mpc$.  (Figure from Ref.~\protect\cite{Strauss95}.)
\label{fig:slice}}
\end{figure}

\begin{figure}[b]
\centerline{\psfig{figure=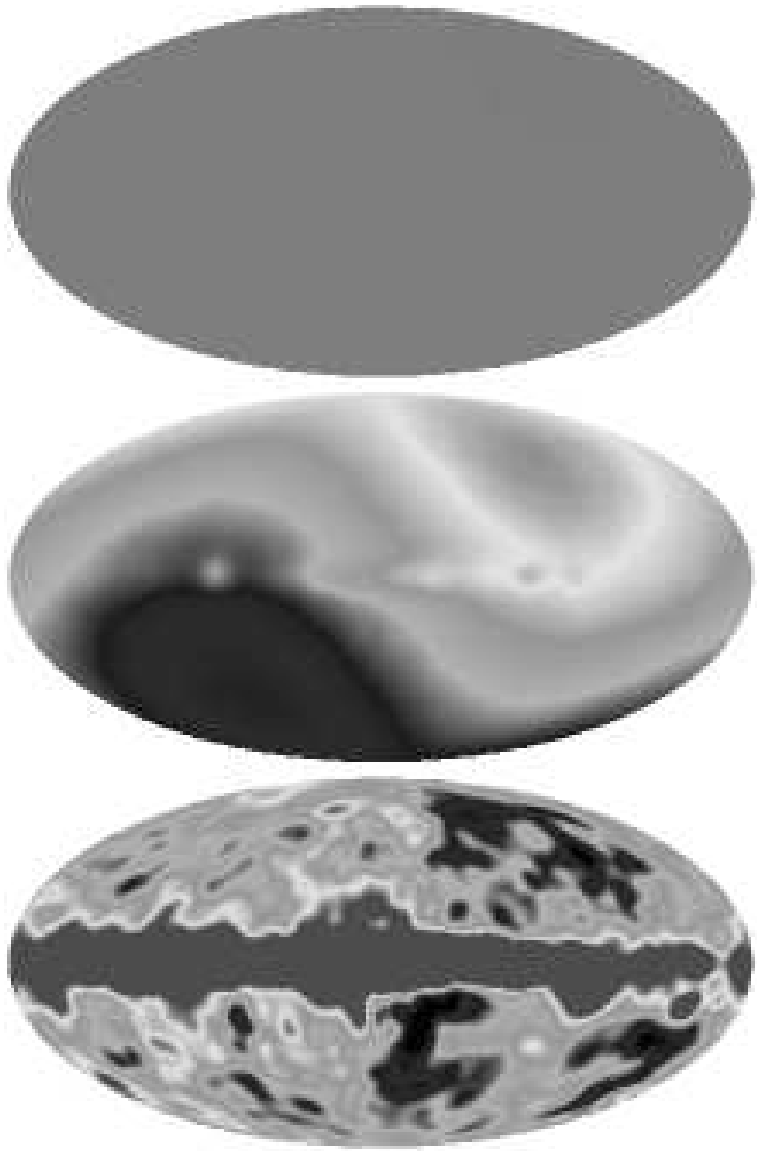,width=10.0cm}}
\caption{Sky map of the temperature of the cosmic microwave background
radiation in the $53\,\rm GHz$ band of the COBE satellite after four
years of data taking~\protect\cite{cobe}.  {\em Top:} The dynamical
range of the color coding is the temperature $T=2.728\,\rm K$ of the
cosmic microwave background, showing its almost perfect isotropy.
{\em Middle:} Dynamical range is $\Delta T =3.353\,\rm mK$ so that the
apparent dipole distribution becomes visible which is attributed to
the Doppler effect from our motion in the cosmic frame of reference.
{\em Bottom:} $\Delta T=18\,\mu\rm K$, showing tiny temperature
fluctuations on all angular scales down to the resolution of the
experiment. The
galaxy occupies the horizontal central band; only the temperature
fluctuations at sufficiently large galactic latitudes can be
attributed to the cosmic microwave background rather than to
galactic foreground emission. 
\label{fig:cobe}}
\end{figure}

Very dramatic constraints on the nature of dark matter arise from
arguments of cosmic structure formation. At early times when the
cosmic microwave background (CMB) radiation decoupled from the ambient
plasma the universe was extremely smooth as demonstrated by the tiny
amplitude of the temperature fluctuations of the CMB across the sky
(Fig.~\ref{fig:cobe}). The present-day distribution of matter, on the
other hand, is very clumpy. There are stars, galaxies, clusters of
galaxies, and large-scale coherent structures on scales up to about
$100\,\rm Mpc$. This is evident, for example, from the density map of
Fig.~\ref{fig:dekel} and also directly apparent from galaxy redshift
surveys~\cite{Strauss95,Strauss96}. The ``stick man'' of the CfA
redshift survey (Fig.~\ref{fig:slice}) has become an icon for
structure in the large-scale galaxy distribution. This picture shows
the distribution of galaxy redshifts along a strip in the sky.
Because redshifts measure distance through Hubble's law (apart from
the peculiar-motion component which cannot be removed without an
independent distance indicator) this sort of representation gives one
a direct visual impression of the three-dimensional galaxy
distribution. A similar picture from the Las Campanas Redshift Survey
is shown in Fig.~\ref{fig:bigslice} which goes out to much larger
distances and thus demonstrates that there is large-scale structure,
but also that there do not seem to be coherent structures on scales
even larger than about $100\,\rm Mpc$. Therefore, it appears justified
to think of the universe as homogeneous on the largest scales.

\begin{figure}[b]
%\centerline{\psfig{figure=fig16.eps,width=12.5cm}}
\vskip16.5cm
\caption{The galaxy distribution from the Las Campanas Redshift
Survey. The three slices in the northern and southern galactic caps
are each shown projected on top of one another.
(Figure courtesy of Huan Lin.)
\label{fig:bigslice}}
\end{figure}

A perfectly homogeneous expanding universe stays that way forever;
there will be no structures.  The standard theory~\cite{Boerner,
  KolbTurner, Coles, Primack} for the formation of structure assumes
that the universe was initially almost, but not quite, perfectly
homogeneous, with a tiny modulation of its density field. The action
of gravity then works to enhance the density contrast as time goes on,
leading to the formation of galaxies or clusters when the self-gravity
of an overdense region becomes large enough to decouple itself from
the overall Hubble expansion. Larger structures have not yet ``gone
nonlinear'' in this sense, yet the distribution of matter shows the
result of the gravitational re-arrangement of the original
distribution.

The outcome of this evolution depends on the initial spectrum of
density fluctuations. The power spectrum of the primordial density
field is usually taken to be approximately flat, i.e.\ of the
``Harrison-Zeldovich-type,'' which corresponds to the power-law-index
$n=1$. However, the {\em effective\/} spectrum relevant for structure
formation is the processed spectrum which obtains at the epoch when
the universe becomes matter dominated as it is only then that
fluctuations can begin to grow by the gravitational instability
mechanism. Because the matter which makes up the cosmic fluid can
diffuse, the smallest-scale primordial density fluctuations will be
wiped out. This effect is particularly important if the density is
dominated by weakly interacting particles which can diffuse far while
they are relativistic. Low-mass particles stay relativistic for a long
time and thus wipe out the primordial fluctuation spectrum up to large
scales.  Massive particles stay put earlier and thus have this effect
only on small scales. One speaks of ``hot dark matter'' (HDM) if the
particle masses are small enough that all fluctuations are wiped out
beyond scales which later correspond to a galaxy. Conversely, ``cold
dark matter'' (CDM) has this effect only on sub-galactic scales. In
the CDM picture smallest structures form first (bottom-up) while in
HDM large structures form first and later fragment into smaller ones
such as galaxies (top-down).

\begin{figure}
\centerline{\psfig{figure=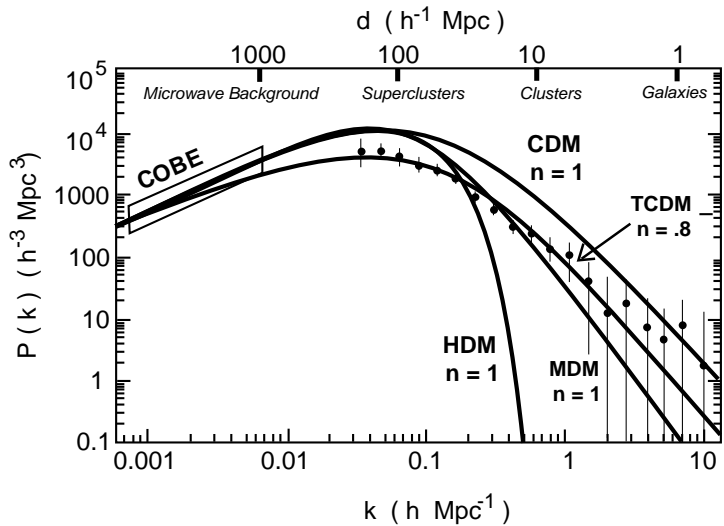,width=10cm}}
\caption{Comparison of matter-density power spectra for cold dark
matter (CDM), tilted cold dark matter (TCDM), hot dark matter (HDM),
and mixed hot plus cold dark matter (MDM) for large-scale structure
formation \protect\cite{Steinhardt}. All curves are
normalized to COBE and include only linear approximation; nonlinear
corrections become important on scales below about
$10\,\rm Mpc$.
\label{fig:struc}}
\end{figure}

One way of presenting the results of calculations of structure
formation is to show the expected power-spectrum of the present-day
matter distribution (Fig.~\ref{fig:struc}) which can be compared to
the observed galaxy distribution. The theory of structure formation
then predicts the form, but not the amplitude of the spectrum which
can be fit either on large scales to the observed temperature
fluctuations of the cosmic microwave background as observed by the
COBE satellite, or else on small scales to the observed galaxy
distribution. Fig.~\ref{fig:struc} illustrates that hot dark matter
(low-mass neutrinos) suppresses essentially all small-scale structure
below a cut-off corresponding to a supercluster scale and thus does
not seem to be able to account for the observations. While cold dark
matter works impressively well, it has the problem of producing too
much clustering on small scales. Ways out include a primordial power
spectrum which is not strictly flat (tilted dark matter), a mix of
cold and hot dark matter, or the assumption of a cosmological
constant.

All of this leaves the question open where the primordial density
fluctuations came from. The standard answer is provided by the
inflationary-universe scenario which traces the density fluctuations
to quantum fluctuations in the very early universe which were boosted
to macroscopic scales during a phase of exponential expansion
(inflation).

It is also possible that the original ``seeds'' for structure
formation are not density fluctuations of the primordial medium, but
rather topological defects from a primordial phase transition such as
``textures'' or ``cosmic strings'' \cite{Shellard,Avelino}.  However,
such scenarios are now widely disfavored because the simplest
gravitational instability picture works so well and because cosmic
microwave observations already seem to rule out at least some variants
of these theories~\cite{Pen}.

Of course, the most important dark-matter question is if a purely
baryonic universe is possible. Standard big-bang nucleosynthesis
already negates this option, and structure formation yields further
counter arguments. A primordial fluctuation spectrum in baryons
consistent with the COBE measurements does not allow the observed
structure to form until today. Weakly interacting particles fare
better because they can begin to form structure earlier than baryonic
matter which is held up by photon pressure until the time of
decoupling (``dark matter boost''). One can circumvent this argument
by preventing baryon density fluctuations from imprinting themselves
on the cosmic microwave background (``isocurvature fluctuations''),
leading to ``primordial isocurvature baryon'' (PIB) scenarios. In view
of the current cosmic microwave background observations, however, such
scenarios seem to be essentially excluded~\cite{Hu}.

Currently there is a wide consensus that some variant of a CDM
cosmology where structure forms by gravitational instability from
a primordial density fluctuations of the Harrison-Zeldovich type
is probably how our universe works. Which does not mean, of course,
that one should prematurely discard possible alternatives such as
cosmic-string induced structure formation.

\subsection{Cosmic Microwave Background}
\label{sec:CosmicMicrowaveBackground}

The cosmic microwave background radiation holds such a wealth of
information that it has been rightly termed ``The Cosmic Rosetta
Stone'' \cite{cmb}. Its very presence in the universe and its
uncannily precise black-body nature are the most striking proofs of
the hot big-bang cosmogony. The observation of tiny angular
temperature variations (Fig.~\ref{fig:cobe}) with typically
$10\,\mu\rm K$ amplitudes already provides tight constraints on
theories of structure formation as outlined in the previous section.
Since the first full-sky COBE maps appeared, a wealth of information
from ground-based and balloon-borne measurements on smaller angular
scales for patches of the sky have become available~\cite{Smoot}.
 
\begin{figure}[b]
\centerline{\psfig{figure=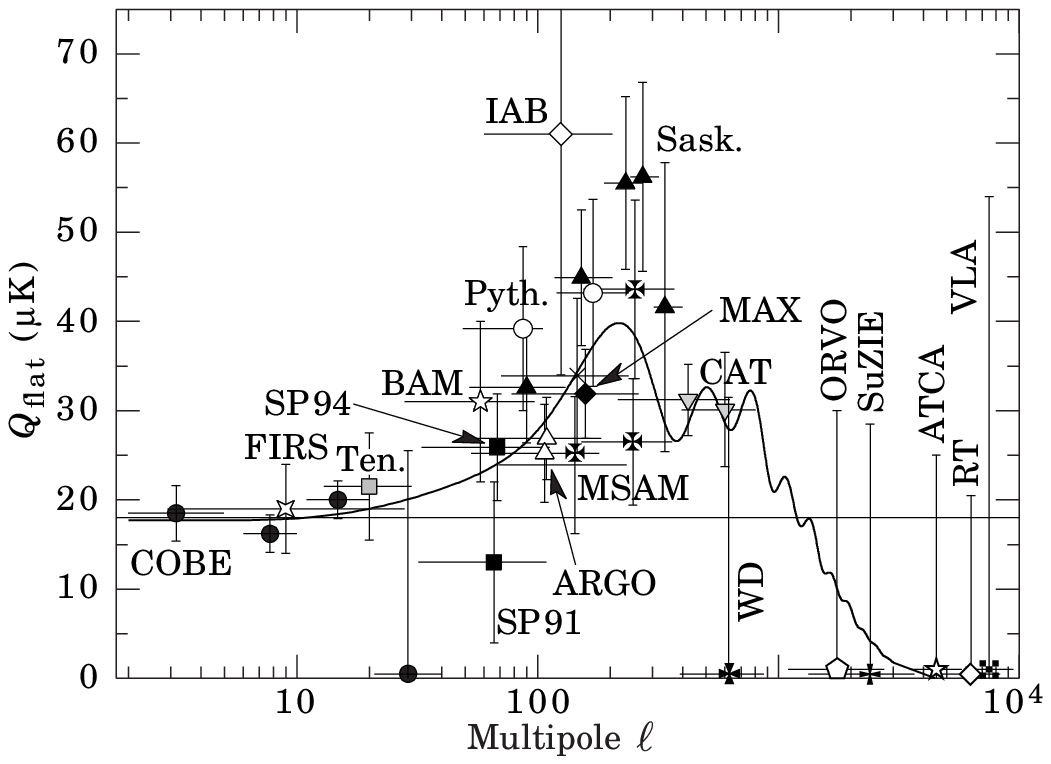,width=8.0cm}}
\vskip0.5cm
\centerline{\psfig{figure=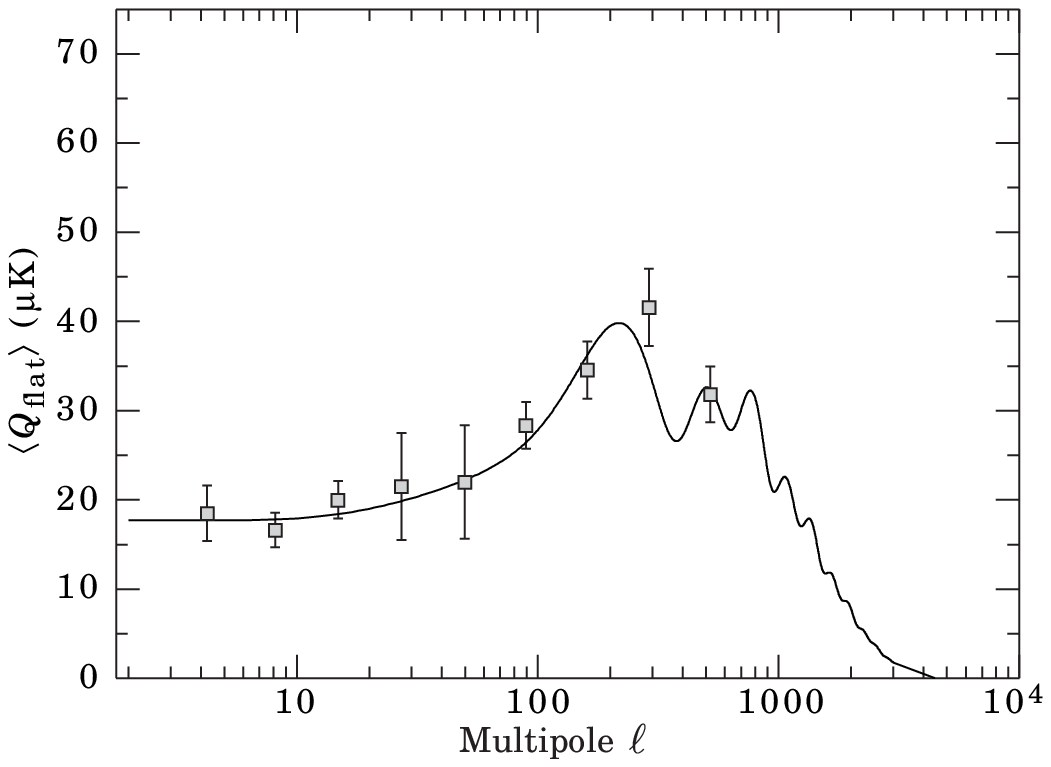,width=8.0cm}}
\caption{Current status of CMB anisotropy observations from COBE and a
large number of ground-based or balloon-borne
experiments~\protect\cite{Smoot}.  Plotted is the equivalent
quadrupole moment of a sky map with a flat power spectrum of
temperature fluctuations which provides the experimentally measured
point at the angular scale of a given experiment. The theoretical
curve corresponds to a standard CDM model with $\Omega_{\rm M}=1$,
$\Omega_{\rm B}=0.05$, and $h=0.5$. The lower panel shows combined
results for a number of angular bins. While this is not a
statistically rigorous procedure it gives a good impression of the
overall trend.
\label{fig:cmb}}
\end{figure}

The important cosmological information is not contained in a
coordinate-space sky map as in Fig.~\ref{fig:cobe} but rather in its
statistical properties. One usually considers the power spectrum of
the temperature map in a spherical-harmonic expansion. Because of the
overall cosmic isotropy one sums over all $m$ for a given multipole
order $\ell$ and analyses the power spectrum as a function of
$\ell$. In Fig.~\ref{fig:cmb} the current data are shown together with
theoretical predictions for a standard CDM universe.  While there is
still a lot of scatter in the data, they already seem to confirm the
appearance of the first ``Doppler peak.'' If structure forms by
initial seeds such as textures or cosmic strings the predicted
spectrum would not show acoustic peaks.  Therefore, at least some
variants of such models already seem to be ruled out~\cite{Pen}.

\begin{figure}[b]
\centerline{\psfig{figure=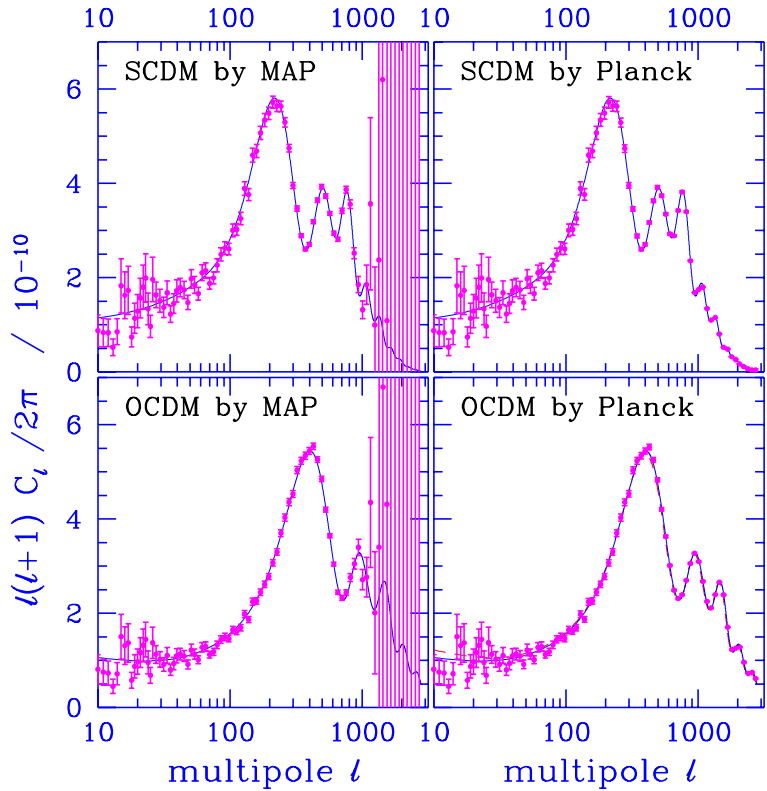,width=12.0cm}}
\caption{Temperature power spectra for a standard CDM (SCDM) model
with $\Omega_{\rm M}=1$ and $h=0.5$, and an open model (OCDM) with
$\Omega_{\rm M}=0.33$ and $h=0.6$, each time without a cosmological
constant~\protect\cite{Tegmark}. Also shown are examples of expected
data sets for simulated CMB sky maps.
\label{fig:max}}
\end{figure}

The predicted pattern of the ``acoustic peaks'' in the power spectrum
is a direct manifestation of rather fine details of the cosmological
model~\cite{cmbreviews}.  In Fig.~\ref{fig:max} the theoretical
predictions are shown for a standard and an open CDM model. One can
easily see how conspicuously the pattern is shifted between the two
models.  There are two approved satellite missions, NASA's Millimeter
Anisotropy Probe (MAP) to be launched in 2000 and ESA's Planck
Surveyor to be launched around 2004 which will take full-sky
temperature maps at much smaller angular resolutions than COBE could
do.  Fig.~\ref{fig:max} also shows a set of simulated measurement
results for both MAP and Planck if the SCDM or OCDM models happen to
represent our universe. Evidently the cosmological parameters can be
pinned down with great precision. It is thought that these experiments
will be able to determine the most important cosmological parameters
eventually on the 1\% level, notably the baryon fraction and total
dark matter content~\cite{CMBparameters}. There remain degeneracies
between different combinations of parameters, however, which will need
to be broken by other methods.

%%%%%%%%%%%%%%%%%%%%%%%%%%%%%%%%%%%%%%%%%%%%%%%%%%%%%%%%%%%%%%%%%%%%%%
%% Section IV %%%%%%%%%%%%%%%%%%%%%%%%%%%%%%%%%%%%%%%%%%%%%%%%%%%%%%%%
%%%%%%%%%%%%%%%%%%%%%%%%%%%%%%%%%%%%%%%%%%%%%%%%%%%%%%%%%%%%%%%%%%%%%%

\section{CANDIDATES AND SEARCHES}

\subsection{Dark Stars (MACHOs)}

The existence of huge amounts of dark matter in the universe at large
and in our own galaxy in particular is now established beyond any
reasonable doubt, but its physical nature remains an unresolved
mystery. A number of compelling arguments relating to big-bang
nucleosynthesis, the amount of x-ray gas in galaxy clusters, and the
small CMB anisotropies in conjunction with theoretical
structure-formation arguments negate the possibility of a purely
baryonic universe. However, there is a big difference between
compelling yet circumstantial arguments and a direct proof. Therefore,
one may still ask if the galactic dark halo could at all consist of
purely baryonic material in some nonluminous form, and if so, how one
should go about to detect~it. Moreover, the same arguments which
indicate that the universe is not purely baryonic motivate significant
amounts of dark baryons which must be hiding somewhere.

\begin{figure}[b]
\centerline{\psfig{figure=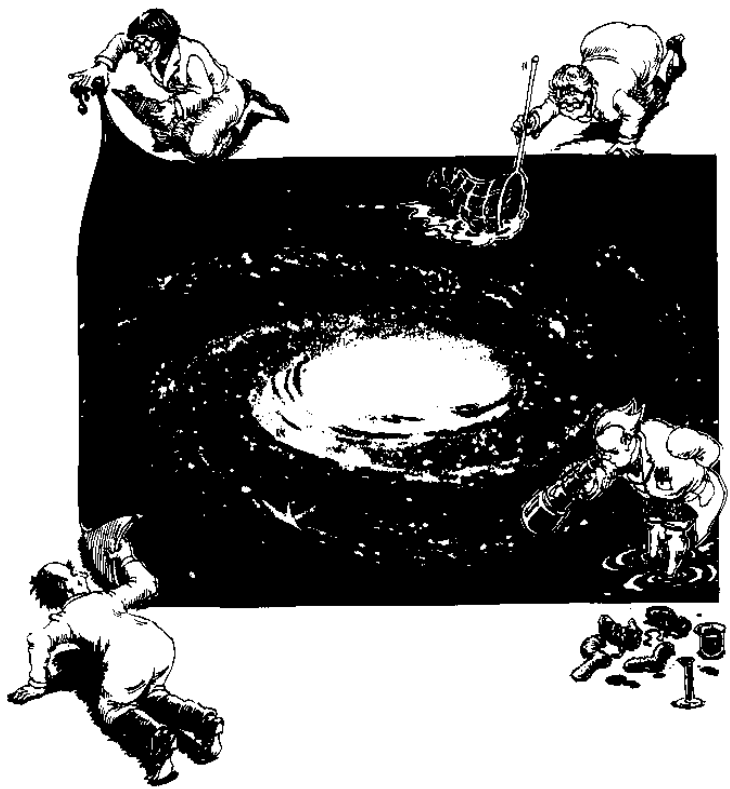,width=10.0 cm}}
\caption{The search for dark matter in the Milky Way.
Reprinted with permission of David Simonds (c). 
\label{fig:eco}}
\end{figure}

Assuming there are dark baryons in the galactic halo, which form could
they take? Evidently they are not in the form of normal and thus
luminous stars or in the form of hot (and thus shining) or cold (and
thus absorbing) gas or dust. In terms of stellar objects one is thus
left with stars which are too small to shine brightly (brown dwarfs or
M-dwarfs) or with burnt-out stellar remnants (white dwarfs, neutron
stars, black holes). Stellar remnants seem implausible because they
arise from a population of normal stars of which there is no trace in
the halo. Neutron stars or black holes, in particular, typically would
form in supernova explosions of which there cannot have been too many
in the galaxy without contaminating it with ``metals,'' i.e.\ elements
heavier than hydrogen and helium. The overproduction of helium also
constrains the presence of white dwarfs which are remnants of stars
not massive enough to reach the supernova phase. White dwarfs as a
dominant halo component cannot be rigorously excluded.  However,
besides the problem of the helium overproduction they would require an
extremely special stellar initial mass function (IMF) with masses
strongly peaked between 2 and $8\,M_\odot$ (solar mass) to avoid the
overproduction of supernovae (for heavier masses) and of normal stars
which would still shine today. These sort of arguments are explained
in more detail in Refs.~\cite{Carr,Hegyi}.

For small stars one distinguishes between M-dwarfs with a mass below
about $0.1\,M_\odot$ which are intrinsically dim and brown dwarfs with
$M\alt 0.08\,M_\odot$ which never ignite hydrogen and thus shine even
more dimly from the residual energy due to gravitational
contraction. The stellar mass function rises towards small masses
(most stars are small) so that one expects significant numbers of such
obejcts in the galaxy.  However, if the galactic halo were to consist
of dim stars would leave one wondering why this population contains so
few higher-mass and thus luminous stars which form so easily in the
disk.  In any event, very long-exposure images of the Hubble Space
Telescope restrict the possible M-dwarf contribution of the galactic
halo to below~6\%~\cite{HSTlimit}.

An extrapolation of the stellar mass function to small masses predicts
large numbers of brown dwarfs within normal stellar populations, but
their very existence has been difficult to prove~\cite{Stevenson} with
only one firm candidate now established~\cite{browndwarf}. While the
paucity of luminous stars in the galactic halo argues against brown
dwarfs, they are the most plausible baryonic candidate for the
galactic dark matter~\cite{Carr}. This conclusion can be avoided if
the halo is very clumpy, allowing for the possibility of
gravitationally bound clouds of molecular hydrogen which are very
difficult to detect and perhaps clumps of dim stars~\cite{Jetzer}.

\begin{figure}[b]
\centerline{\psfig{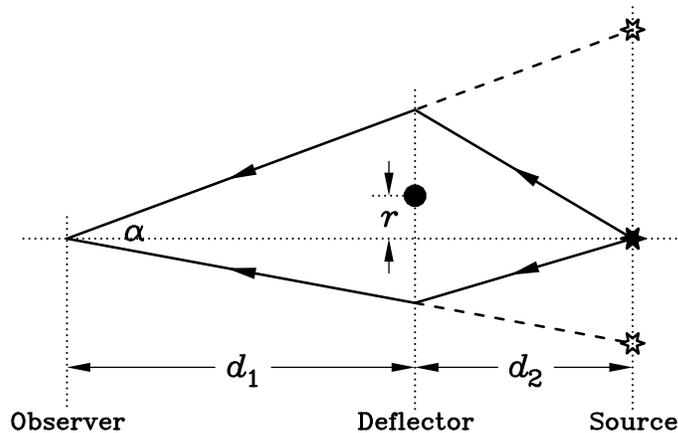}}
\caption{Geometry of light deflection by a pointlike mass which yields
two images of a source viewed by an observer. 
\label{fig:micro}}
\end{figure}

Whatever the merits of the arguments for or against baryonic objects
as galactic dark matter, nothing would be more convincing than a
direct detection of the candidates or their exclusion in a search
experiment. Fortunately, in 1986 Paczy\'nski proposed an exciting
method to search systematically for faint stars in the halo of our
galaxy~\cite{Paczynski}. His idea is based on the well-known
effect~\cite{Chwolson} that a ``pointlike'' mass (deflector) placed
between an observer and a light source creates two distinct images as
indicated in Fig.~\ref{fig:micro}. (A nonsingular and transparent mass
distribution always yields an odd number of images.) When the source
is exactly aligned behind the deflector (mass $M_{\rm D}$) the image
would be an annulus instead (``Einstein ring'') with a radius
(``Einstein radius'') of
\begin{equation}
r_{\rm E}=\sqrt{G_{\rm N} M_{\rm D} d}
\hbox{\qquad where\qquad}
d\equiv \frac{4 d_1 d_2}{d_1+d_2}
\end{equation}
with the distances $d_{1,2}$ as in Fig.~\ref{fig:micro}.  Because of
differential bending of the ``rays'' which produce the images, the
image brightnesses will be different from each other and from the
single image in the absence of gravitational lensing. If the two
images cannot be separated because their angular distance $\alpha$ is
below the resolving power of the observer's telescope, the only effect
will be an apparent brightening of the star, an effect known as
``gravitational microlensing.'' The magnification (``amplification'')
factor is
\begin{equation}
A=\frac{2+u^2}{u\sqrt{4+u^2}}
\hbox{\qquad where\qquad} u\equiv \frac{r}{r_{\rm E}}
\end{equation}
and $r$ is the distance of the deflector from the line of sight. 

\begin{figure}[ht]
\centerline{\psfig{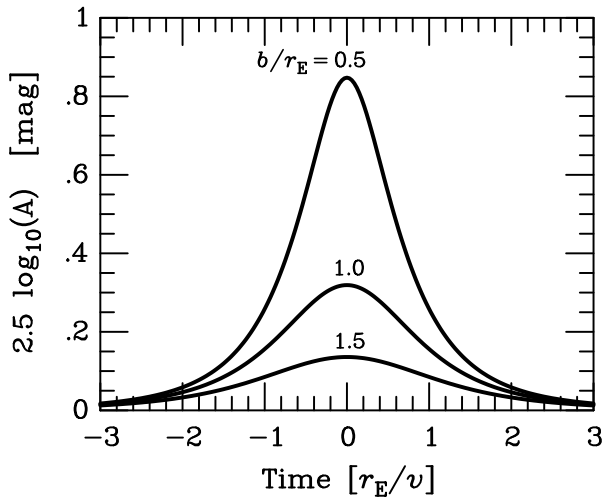}}
\caption{Apparent lightcurve of a source if a pointlike deflector
passes the line of sight with a transverse velocity $v$ and an
``impact parameter'' $b$. The vertical axis for the magnification
(``amplification'') factor $A$ was chosen logarithmically and
multiplied with 2.5 to obtain the usual astronomical logarithmic
brightness measure ``magnitude'' or mag.
\label{fig:ampli}}
\end{figure}

If we imagine a terrestrial observer watching a distant star, and
if the galactic halo is filled with ``massive astrophysical halo
objects'' (MACHOs), one of them will occasionally pass near the line
of sight and thus cause the image of the monitored star to
brighten. If the deflector moves with the velocity $v$ transverse to
the line of sight, and if its ``impact parameter'' (minimal distance
to the line of sight) is $b$, then one expects an apparent lightcurve
as shown in Fig.~\ref{fig:ampli} for several values of $b/r_{\rm
E}$. The natural ``unit of time'' is $r_{\rm E}/v$, the origin was
chosen at the time of closest approach to the line of sight.

A convenient sample of target stars is provided by the Large
Magellanic Cloud (LMC) which is a small satellite galaxy of the Milky
Way at a distance from us of about $50\,\rm kpc$. It has enough bright
stars, it is far enough away so that the line of sight intersects a
significant fraction of the galactic halo, and it is far enough above
the galactic plane so that one actually cuts through the halo, not
just through the galactic disk. Any given star in the LMC will be
substantially brightened at the time of observation if the line of
sight intersects with the circular cross section $\pi r_{\rm E}^2$
around some MACHO. If the halo is supposed to be made of such objects,
their number density is inversely proportional to their assumed mass
while $\pi r_{\rm E}^2$ is directly proportional to it. Therefore, the
probability for a target star to be lensed at the instance of
observation is independent of the mass of the dark-matter objects. For
stars in the LMC one finds a probability (``optical depth for
microlensing of the galactic halo'') of $\sim 10^{-6}$. Put another
way, if one looks simultaneously at $\sim 10^{6}$ stars in the LMC one
has a good chance of seeing at least one of them brightened by a dark
halo star.

\begin{figure}[b]
\centerline{\psfig{figure=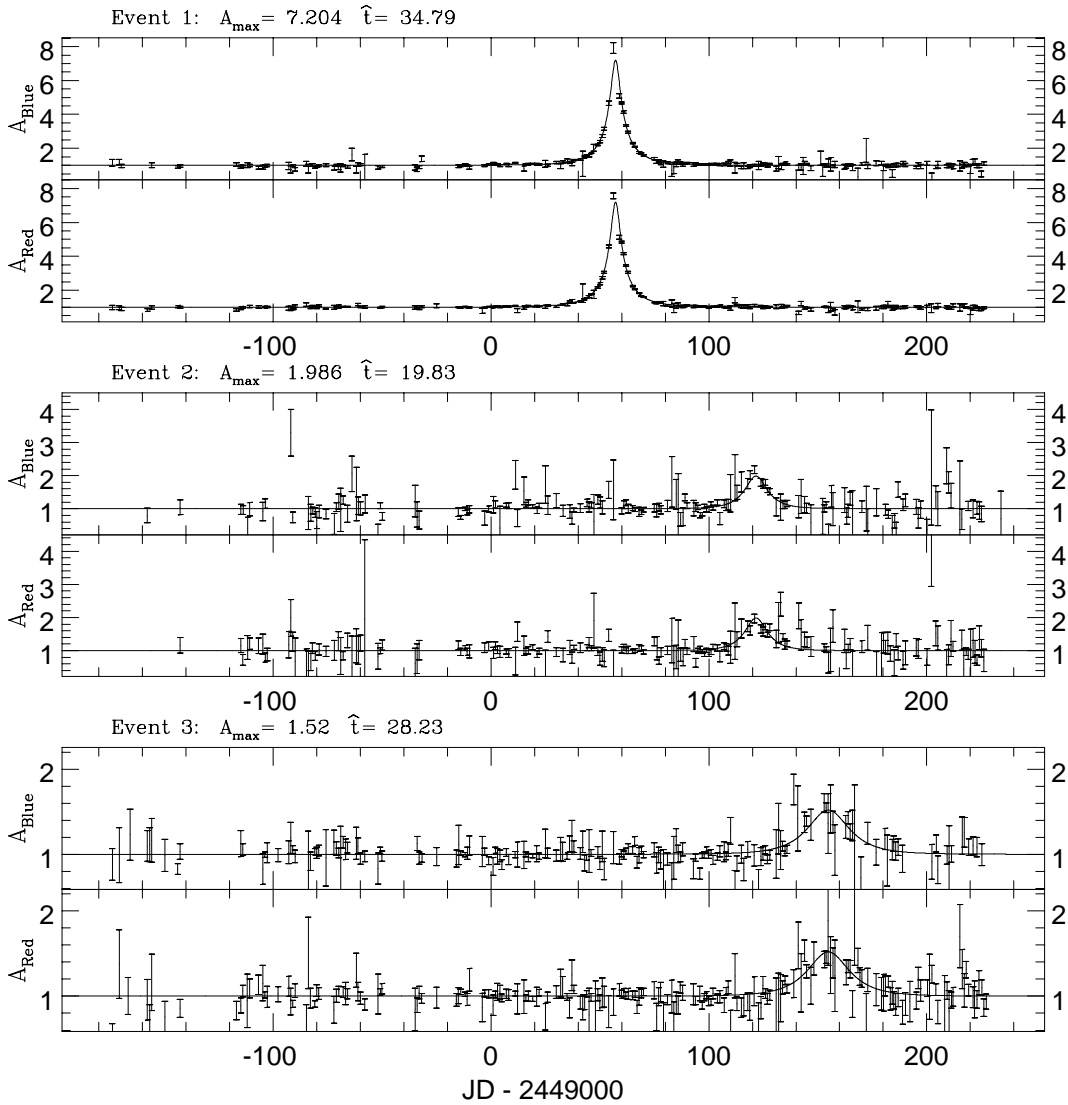,width=14.0cm}}
\caption{Red and blue lightcurves (amplification factors)
  of the first three microlensing candidates of the MACHO
  collaboration toward the LMC~\protect\cite{Macho95}. The horizontal
  axis (time) is measured in days with the zero point at 
  Julian Day 2449000, corresponding to January 12, 1993.
\label{fig:macho}}
\end{figure}

In order to recognize a lensing event one has to monitor this large
sample of stars long enough to identify the characteristic lightcurve
shown in Fig~\ref{fig:ampli}. It has the property of being unique,
symmetric about $t=0$, and achromatic, three signatures which allow
one to discriminate against normal variables which comprise about 1\%
of all stars. The typical duration of the apparent brightness
excursion is $r_{\rm E}/v$, i.e.\ the time it takes a MACHO to cross
an Einstein radius, which depends on the MACHO mass. If the deflector
mass is $1\,M_\odot$ (solar mass) a mean microlensing time will be 3
months, for $10^{-2}\,M_\odot$ it is 9 days, for $10^{-4}\,M_\odot$ it
is 1 day, and for $10^{-6}\,M_\odot$ it is 2~hours.

The microlensing search for dark stars was taken up by the MACHO and
the EROS collaborations, both reporting first tentative candidates
toward the LMC in 1993 \cite{firstmachos}; the lightcurves for the
first three MACHO candidates are shown in
Fig.~\ref{fig:macho}. Because one did not expect the galactic halo to
consist dominantly of dark stars these findings were quite sensational
at the time. Meanwhile, more candidates have appeared, perhaps a dozen
or so toward the LMC. Moreover, the galactic bulge has been used as
another target where many more events occur through microlensing by
ordinary disk stars. While observations of the galactic bulge are not
sensitive to halo dark-matter stars, they allow one to develop a good
understanding of the microlensing technique itself, and anyhow are
interesting as a method to study the structure of the galactic bulge
and disk and their stellar content. It is now established beyond doubt
that the microlensing technique works. Within the past few years it
has established itself as a completely new approach to galactic
astronomy, with at least half a dozen collaborations pursuing
observations of various target regions. As a by-product these searches
naturally produce a huge database of intrinsically variable stars
which is invaluable to stellar astronomy, independently of the
dark-matter problem.

\begin{figure}[b]
\centerline{\psfig{figure=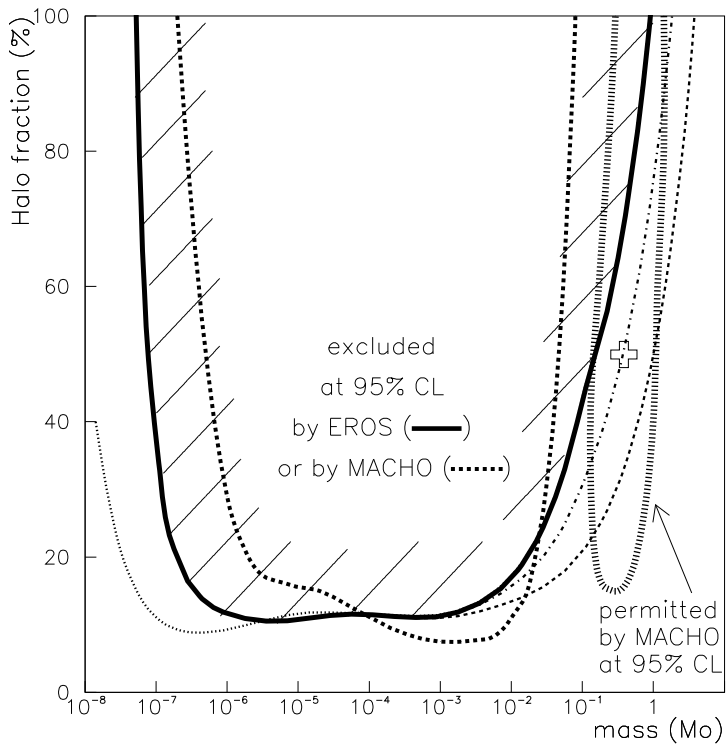,width=10.0cm}}
\caption{Exclusion diagram at 95\% C.L.\ for the halo fraction and
mass of the assumed MACHOs~\protect\cite{EROS97}.  It was assumed that
they all have the same mass and that a standard model for the galactic
halo obtains. The dotted line on the left is the limit when blending
and finite size effects are ignored in the EROS limits. The dot-dashed
and dotted lines on the right are the EROS limits when 1 or 2 of their
candidate events are attributed to MACHOs. The cross is centered on
the 95\% C.L.\ permitted range of the MACHO collaboration for their
case of a standard spherical halo~\protect\cite{Macho97}.
\label{fig:eros}}
\end{figure}

Far from clarifying the status of dim stars as a galactic dark matter
contribution, the results of the current microlensing results toward
the LMC are quite confusing~\cite{Macho95,EROS95,EROS97,Macho97}. If
one assumes a standard spherical galactic halo the absence of
short-duration events excludes a large range of MACHO masses as a
dominant halo component (Fig.~\ref{fig:eros}).  On the other hand,
assuming all MACHOs have the same mass one finds a best-fit mass of
about $0.4\,M_\odot$ and a halo fraction which could be anything
between about 10\% and 100\% (Fig.~\ref{fig:eros}). The best-fit mass
is characteristic of white dwarfs, but a galactic halo consisting
primarily of white dwarfs is highly implausible and barely compatible
with a variety of observational constraints.  On the other hand, if
one wanted to attribute the observed events to brown dwarfs ($M\alt
0.08\,M_\odot$) one needs to appeal to a very nonstandard density
and/or velocity distribution of these objects. Other explanations
involve an unexpectedly large lensing contribution from LMC stars, a
thick galactic disk contribution, an unrecognized population of normal
stars on the line of sight to the LMC, and other speculations, with
pros and cons for each hypothesis~\cite{LMCexplanations}. At the
present time it is absolutely unclear which sort of objects the
microlensing experiments are seeing toward the LMC and where the
lenses are.

Meanwhile a first event has appeared in both the MACHO
and EROS data toward the Small Magellanic Cloud (SMC) \cite{SMC},
another galactic satellite at a slightly larger distance than the LMC
and about $20^\circ$ away in the sky. While one event does not carry
much statistical significance, its appearance is consistent with the
LMC data if they are interpreted as evidence for halo dark matter.
However, this interpretation would imply a large mass (a few solar
masses) for the lens due to the large duration. 

Besides more data from the LMC and SMC directions, other lines of
sight might provide significant information on the stellar make-up of
the galactic halo. Of particular importance is the Andromeda galaxy
(Fig.~\ref{fig:M31}) as a target because the line of sight cuts
through the halo almost vertically relative to the galactic disk.
Unfortunately, Andromeda is so far away that one cannot resolve
individual target stars for the microlensing purpose. One depends on
the ``pixel lensing'' technique where one observes the apparent
brightening of a single pixel of the CCD camera; one pixel covers the
unresolved images of many stars. At least two groups pursue this
approach which already has produced preliminary limits~\cite{pixel}.

\subsection{Neutrinos}

In spite of the puzzling observation of microlensing events toward the
Large and Small Magellanic Clouds which may indicate that some of the
galactic dark matter is in the form of dim stars, the case for a
dominant dark-matter component in the form of weakly interacting
particles is rather compelling.  A purely baryonic universe is at odds
with the baryon fraction implied by big-bang nucleosynthesis and the
amount of x-ray gas in galaxy clusters. Most importantly, the
formation of structure by the generic gravitational instability
mechanism does not work with baryons alone, while it is impressively
successful with weakly interacting particles as dark~matter.

\begin{figure}
\centerline{\psfig{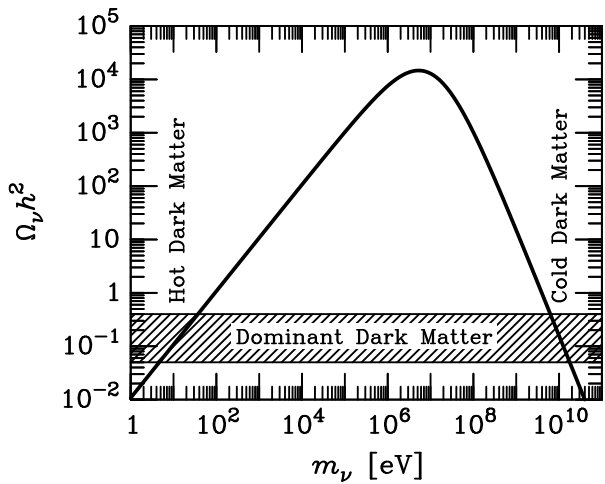}}
\caption{Cosmic neutrino mass density as a function of neutrino
mass. The hatched band indicates the range for $\Omega h^2$ which the
dominant particle dark matter component must provide according to
Eq.~(\protect\ref{eq:pdm}).
\label{fig:leew}}
\end{figure}

The only candidates which are currently known to exist are
neutrinos. In order to understand if they could represent the dark
matter we need to calculate their cosmic abundance as a function of
their assumed mass. If it is small this is a straightforward exercise;
in the framework of the hot big-bang cosmogony one expects about as
many cosmic ``black-body neutrinos'' as there are microwave
photons. In detail, the cosmic energy density in massive neutrinos is
found to be $\rho_\nu=\frac{3}{11}\,n_\gamma\,\sum m_\nu$ with
$n_\gamma$ the present-day density in microwave background
photons~\cite{KolbTurner}.  The sum extends over the masses of all
sequential neutrino flavors. In units of the critical density this is
(Fig.~\ref{fig:leew})
\begin{equation}
\Omega_\nu h^2=\sum \frac{m_\nu}{93\,\rm eV}.
\end{equation}
The observed age of the universe together with
the measured expansion rate yields $\Omega h^2\alt 0.4$ so that for
any of the three families
\begin{equation}
m_\nu\alt 40\,{\rm eV}.
\end{equation}
This mass limit is probably the most important astrophysical
contribution to neutrino physics because for $\nu_\tau$ it improves
the experimental limit by about six orders of magnitude.

It is also interesting to ask for a lower limit on $\Omega h^2$ which
the dominant dark-matter component must obey. According to
Eq.~(\ref{eq:flowmass}) the matter density is limited by $\Omega_{\rm
M}\agt0.3$.  Allowing for a significant baryon fraction indicates that
particle dark matter (PDM) should obey $\Omega_{\rm
PDM}\agt0.2$. Taking $h\agt0.5$ as a lower limit for the expansion
rate implies
\begin{equation}\label{eq:pdm}
0.05\alt \Omega_{\rm PDM}h^2\alt0.4
\end{equation}
as a reasonable range where a given particle dark matter candidate
could be all of the nonbaryonic dark matter (hatched band in
Fig.~\ref{fig:leew}).  Therefore, neutrinos with a mass $4\,{\rm
eV}\alt m_\nu\alt40\,{\rm eV}$ could represent all of the nonbaryonic
dark matter.

There is a second solution at large masses.  If the mass significantly
exceeds the cosmic temperature at a given epoch, the neutrino density
is suppressed by a Boltzmann factor $e^{-m_\nu/T}$.  The weak
interaction rates in the early universe become slow relative to the
overall expansion when the temperature falls below about $1\,\rm MeV$.
For masses exceeding this weak freeze-out temperature the Boltzmann
suppression occurs while the neutrinos are still in thermal
equilibrium, reducing the relic density accordingly.  A detailed
calculation of the relic density requires an approximate solution of
the Boltzmann collision equation~\cite{KolbTurner}. Apart from a
logarithmic correction one finds $\Omega_\nu h^2\propto m_\nu^{-2}$ as
shown in Fig.~\ref{fig:leew} for the Majorana case.  Dirac neutrinos
have a slightly smaller relic density, but in either case neutrinos
could be the dark matter if their mass was a few GeV. The laboratory
limit for $\nu_\tau$ of about $20\,\rm MeV$, and more restrictive ones
for $\nu_\mu$ and $\nu_e$, precludes this possibility among the known
sequential~flavors.

Low-mass neutrinos, however, are problematic dark matter
candidates from the perspective of structure formation because they
represent ``hot dark matter'' (Sec.~\ref{sec:structureformation}).
Forming small-scale structure such as galaxies would probably require
topological defects such as cosmic strings as seeds for the
gravitational instability, and even then a scenario consistent with
cosmic microwave background constraints may not be possible. 

In addition there is a well-known problem with neutrinos filling the
dark-matter haloes of galaxies. By definition, galactic dark-matter
neutrinos would be gravitationally bound to the galaxy so that their
velocity would be bound from above by the galactic escape velocity
$v_{\rm esc}$, yielding an upper limit on their momentum of $p_{\rm
max}=m_\nu v_{\rm esc}$. Because of the Pauli exclusion principle the
maximum number density of neutrinos is given when they are completely
degenerate with a Fermi momentum $p_{\rm max}$, i.e.\ it is $n_{\rm
max}=p_{\rm max}^3/3\pi^2$. Therefore, the maximum local mass density
in dark-matter neutrinos is $m_\nu n_{\rm max}= m_\nu^4 v_{\rm
esc}^3/3\pi^2$. As this value must exceed a typical galactic dark
matter density, one obtains a {\it lower\/} limit on the necessary
neutrino mass. A refinement of this simple derivation is known as the
Tremaine-Gunn limit~\cite{TremaineGunn}; for typical spiral galaxies
it is about $20\,\rm eV$~\cite{SS97}.

Therefore, dark-matter neutrino masses are squeezed between the upper
limit from the overall cosmic mass density, and the lower limit from
the galactic phase-space argument.  They are squeezed, but perhaps not
entirely squeezed out. Neutrinos could not be the dark matter
of dwarf galaxies where a mass of a few $100\,\rm eV$ is required by
the Tremaine-Gunn argument~\cite{SS97}.  However, perhaps the dark
matter in dwarf galaxies is of a different physical nature. At any
rate, the galactic phase-space argument surely disturbs any
simple-minded fantasy about neutrinos being the dark matter on all
scales.

Neutrinos may still play an important role as dark matter and for
structure formation if they are a subdominant component of a cold-dark
matter (CDM) universe. It was discussed in
Sec.~\ref{sec:structureformation} that CDM produces too much
small-scale structure if the primordial density fluctuation spectrum
was of the Harrison-Zeldovich type. This problem can be patched up by
invoking a mixed hot plus cold dark matter (MDM or CHDM) cosmology
where the hot component erases enough of the initial power on small
scales to compensate for the overproduction by pure CDM~\cite{MDM}.
In a flat universe ($\Omega=1$) the best fit is obtained with a total
mass in neutrinos corresponding to $\sum m_\nu=5\,\rm eV$ with an
equipartition of the masses among the flavors.

The high baryon fraction of galaxy clusters
(Sec.~\ref{sec:xrayclusters}) provides another motivation for a
neutrino dark matter component. If clusters represent a fair sample of
the cosmic matter inventory their high baryon fraction points to a
low-$\Omega$ universe. However, neutrinos are naturally more dispersed
than CDM, providing a less-clustered dark matter background, somewhat
alleviating the cluster baryon problem~\cite{Strickland}.

It will be very difficult to test this hypothesis. A direct detection
of cosmic background neutrinos does not seem to be realistic in the
foreseeable future whether or not they have masses.  It is possible,
of course, that the requisite neutrino mass will appear in neutrino
oscillation experiments; tentative evidence has already been reported
by the LSND Collaboration~\cite{LSND}.  The interpretation of their
$\bar\nu_e$ excess counts in terms of neutrino oscillations implies a
$\nu_e$-$\nu_\mu$ mass difference of order $1\,\rm eV$ or more,
pointing to cosmologically significant neutrino masses. At the present
time one has to wait and see if more LSND data and other experiments,
notably KARMEN~\cite{karmen}, will confirm this claim.

The LSND claim is not easily compatible with the much smaller neutrino
mass differences indicated by the oscillation interpretation of the
solar and atmospheric neutrino anomalies. Of course, oscillation
experiments give us information only about the {\em differences\/} of
neutrino masses, not about their absolute values. Therefore, even if
these differences are small, all neutrinos could have approximately
equal masses with a common offset from zero which could be much larger
than their mass differences. Such scenarios of ``degenerate neutrino
masses'' are not testable by oscillation experiments so that the
direct searches for a $\nu_e$ mass in the eV range by tritium
$\beta$-spectrum experiments and by neutrinoless $\beta\beta$-decay
experiments remain of great importance. Similarly, the observation of
a neutrino signal from a galactic supernova by a detector like
Superkamiokande or the proposed OMNIS~\cite{omnis} would allow one to
detect or exclude sub-eV electron neutrino masses.

\begin{figure}
\centerline{\psfig{figure=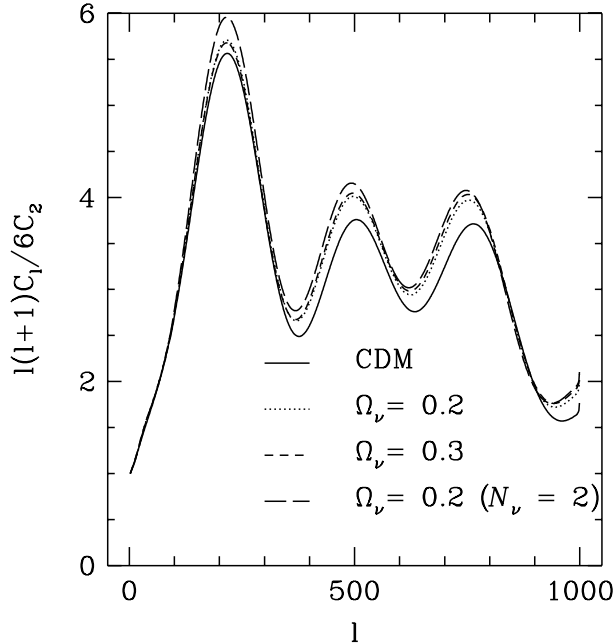,width=8.0cm}}
\caption{Power spectrum of the temperature sky map for the cosmic
microwave background in a cold dark matter cosmology, and three
variants of mixed dark matter~\protect\cite{Stebbins}. 
\label{fig:cmbhot}}
\end{figure}

In Sec.~\ref{sec:CosmicMicrowaveBackground} we discussed the enormous
power of future cosmic microwave background observations to
distinguish between different cosmological models. Could one
distinguish a pure CDM from a CHDM universe? In Fig.~\ref{fig:cmbhot}
the expected power spectrum of the angular temperature fluctuations is
shown for a CDM scenario as a solid line. The modified power spectra
for three versions of CHDM cosmologies are also shown. The resolution
expected from the future microwave satellites is better than the
differences between these curves.  However, there are other unknown
parameters such as the overall mass density, the Hubble constant, the
cosmological constant, the baryon fraction, and so forth, which all
affect the expected power spectrum.  All of them have to be determined
by fitting the power spectrum to the observations, leading to
``degeneracies'' in the sense that not all of these parameters can be
determined independently.  Therefore, it is probably not possible to
identify a small neutrino component by the microwave data
alone. However, in conjunction with the expected precision measurement
of the power spectrum of the matter density from the upcoming Sloan
Digital Sky Survey one would be sensitive to sub-eV neutrino masses,
and even a mass as small as $0.1\,\rm eV$ would make a nonnegligible
difference~\cite{tegmax}.
 
%%%%%%%%%%%%%%%%%%%%%%%%%%%%%%%%%%%%%%%%%%%%%%%%%%%%%%%%%%%%%%%%%%%%%%
%%%%%%%%%%%%%%%%%%%%%%%%%%%%%%%%%%%%%%%%%%%%%%%%%%%%%%%%%%%%%%%%%%%%%%

\subsection{Weakly Interacting Massive Particles (WIMPs)}

Baryons apparently do not make up the bulk of the cosmic matter.
Massive neutrinos are the only alternative among the known particles,
but they are essentially ruled out as a universal dark-matter
candidate, even if they may play an important role as a hot component
in a universe which is otherwise dominated by cold dark matter. What
is the nature of this dominant~component?

From the discussion in the previous section and from
Fig.~\ref{fig:leew} it is apparent that neutrinos with a mass of a few
GeV could well play this role. Their relic abundance would be
appropriate for the cosmic dark matter density, and their large mass
would guarantee that they became nonrelativistic more than early
enough to avoid the erasure of primordial density fluctuations: they
would be cold dark matter. While the experimental mass limits prevent
$\nu_e$, $\nu_\mu$ or $\nu_\tau$ to play this role, a
fourth-generation neutrino was a possibility until the CERN
measurements of the $Z^0$ width showed that there are exactly 3
neutrino families with $m_\nu\alt\frac{1}{2} m_Z=46\,\rm GeV$. With a
mass exceeding this limit the relic abundance would be too low
(Fig.~\ref{fig:leew}).

The calculation of the relic density which leads to the curve of
Fig.~\ref{fig:leew} assumes that heavy neutrinos actually can
annihilate with each other, i.e.\ that there are equal numbers of
neutrinos and antineutrinos in the early universe. For Majorana
neutrinos which are their own antiparticles this represents no extra
constraint, but for Dirac neutrinos this assumption cannot be taken
for granted. After all, the normal matter in the universe survives its
own primordial annihilation only because of the baryon asymmetry of
the universe, and a similar asymmetry could exist in the neutrino
sector. Because of the unknown cosmic asymmetry the relic density of
Dirac neutrinos is not calculable so that one might think that a
fourth-generation Dirac neutrino with a mass beyond the $Z^0$-decay
limit could well play the role of cold dark matter. However, this
possibility is excluded by the direct experimental searches for
galactic dark matter to be discussed below. (As we shall see their
scattering cross section is coherently enhanced so that they are
easier to detect than Majorana neutrinos.)  Either way, a heavy
fourth-generation neutrino would have seemed implausible anyway
because the particles comprising the dark matter must be stable on the
scale of the age of the universe of about $10\,\rm Gyr$. There would
have been no reason to expect a massive fourth-generation neutrino to
be so long-lived. 

However, something like a stable heavy neutrino, a generic {\em Weakly
Interacting Massive Particle (WIMP)}, still seems like a good
possibility because an annihilation cross section given roughly by the
weak scale leaves us with the right relic density and a mass
appropriate for cold dark matter. Naturally, these particles must
couple to $Z^0$ more weakly than sequential neutrinos or else they,
too, would be excluded by the measured $Z^0$ decay width. Fortunately,
supersymmetric extensions of the particle-physics standard model
naturally motivate the existence of the requisite particles in the
form of neutralinos~\cite{JKG96}.

Supersymmetric extensions of the standard model predict a doubling of
the existing particles in that every bosonic degree of freedom is
matched by a supersymmetric fermionic one and vice versa. Normal and
supersymmetric particles differ by a quantum number called R-parity
which may be conserved so that the lightest supersymmetric particle
(LSP) would have to be stable. If the LSP is the lightest
``neutralino,'' i.e.\ the lightest mass eigenstate of a general
superposition of the neutral spin-$\frac{1}{2}$ fermions expected in
this theory, namely the photino (spin-$\frac{1}{2}$ partner of the
photon), Zino (spin-$\frac{1}{2}$ partner of the $Z^0$ boson), and
Higgsino (spin-$\frac{1}{2}$ partner of a neutral Higgs boson), then
we have a perfect ersatz neutrino available. Neutralinos are Majorana
fermions so that their cosmic relic density is determined by the
freeze-out from thermal equilibrium along the lines of
Fig.~\ref{fig:leew} rather than by an unknown cosmic
particle-antiparticle asymmetry. Their interactions would be roughly,
but not exactly, of weak strength. In detail their annihilation and
scattering cross sections depend on specific assumptions about a given
supersymmetric model and on the values of various parameters within
such models.

At the present time no empirical evidence exists that supersymmetric
extensions of the standard model are indeed realized in nature; of
course the search for supersymmetric particles is one of the prime
goals for experiments at future accelerators such as the LHC.  For the
time being the cosmological need for a suitable cold dark matter
candidate is the strongest empirical hint for the reality of the
supersymmetric doubling of the elementary particle zoo.  The
nonobservation of supersymmetric particles at current accelerators
places stringent limits on the neutralino mass and interaction cross
section~\cite{Ellis}.

In the mid-1980s in became clear that even though WIMPs are by
definition weakly interacting particles one can search for them in our
galaxy by a variety of
methods~\cite{JKG96,Goodman,DMReviews,Caldwell90,Caldwell97}.  The
``direct'' approach relies on elastic WIMP collisions with the nuclei
of a suitable target, for example a germanium crystal. Dark-matter
WIMPs move with a typical galactic virial velocity of around $300\,\rm
km\,s^{-1}$. If their mass is $10{-}100\,\rm GeV$ their energy
transfer in such an elastic collision would be of order $10\,\rm
keV$. Therefore, the task at hand is to identify such energy
depositions in a macroscopic sample of a target substance. Of the many
ways that have been discussed to achieve this goal, three are of
particular importance. First, one may search for scintillation light,
for example in NaI crystals or in liquid xenon. Second, one may search
for an ionization signal in a semiconductor, notably in a germanium
crystal. Third, one may cool the target (for example a sapphire
crystal) to very low temperatures of order $10\,\rm mK$ so that a
$10\,\rm keV$ energy deposition causes a measurable temperature
increase. This ``cryogenic'' or ``bolometric'' approach employs a
variety of methods to measure this heating, for example a
superconducting strip attached to the target which is shifted toward
the normal conducting phase by the temperature increase.

The main problem with any such experiment is the extremely low
expected signal rate. In detail it depends on the assumed WIMP
properties and target material, but a typical number is below $1\,\rm
event\,kg^{-1}\,day^{-1}$, a counting-rate unit usually employed in
this field. To reduce natural radioactive contaminations one must use
extremely pure substances and to reduce the background caused by
cosmic rays requires these experiments to be located deeply
underground, for example in the Gran Sasso laboratory or the Boulby
salt mine in England.  All current experiments are essentially
background limited at a level of the order $1\,\rm
event\,kg^{-1}\,day^{-1}$.

Neutrinos scatter on nucleons by virtue of a vector-current and an
axial-vector current (spin-dependent) interaction. For the small
momentum transfers imparted by galactic WIMPs such collisions are
essentially coherent over an entire nucleus, leading to an enhancement
of the effective cross section. The relatively large detection rate
expected in this case allowed one in the late 1980s to exclude
fourth-generation Dirac neutrinos for the galactic dark
matter~\cite{DMReviews}.

However, for Majorana neutrinos the vector-current interaction
vanishes identically; they interact only by a spin-dependent
force. The coherence over the nucleus now works in the opposite
direction: essentially it is the total spin of the nucleus which is
relevant for the scattering rate rather than the scattering rates
summed over the individual nucleons.  Therefore, Majorana neutrinos
are much more difficult to detect. Because neutralinos are of the
Majorana type one largely depends on their spin-dependent interaction
cross section even though they may have a scalar-exchange
contribution, unlike proper neutrinos which interact only by the
exchange of vector bosons.

\begin{figure}[b]
\centerline{\psfig{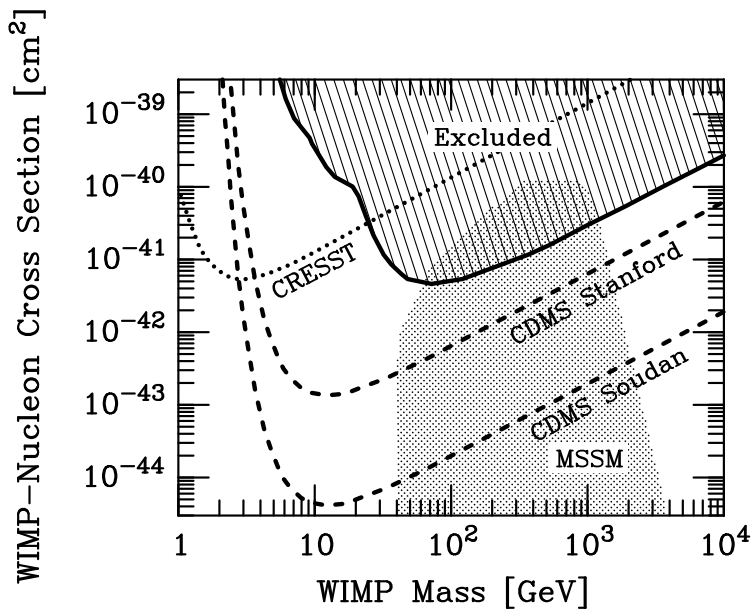}}
\caption{Exclusion range for the spin-independent WIMP scattering
cross section per nucleon from the NaI
experiments~\protect\cite{UKDMC,DAMA96} and the germanium
detectors~\protect\cite{Germanium}. Also shown is the range of
expected counting rates for neutralinos in the minimal supersymmetric
standard model (MSSM) without universal scalar mass
unification~\protect\cite{Gondolo,Bottino}. The search goals for the
upcoming large-scale cryogenic experiments
CRESST~\protect\cite{CRESST} and CDMS~\protect\cite{CDMS} are also
shown, where CDMS is located at a shallow site at Stanford, but will
improve its sensitivity after the planned move to a deep site in the
Soudan mine.
\label{fig:limits}}
\end{figure}

Currently the best limits on WIMP scattering cross sections come from
several germanium experiments~\cite{Germanium}, the NaI scintillation
detectors of the United Kingdom Dark Matter Collaboration (UKDMC)
located in the Boulby salt mine in England~\cite{UKDMC} and of the
DAMA experiment located in the Gran Sasso National Laboratory near
Rome in Italy~\cite{DAMA96}. In Fig.~\ref{fig:limits} the current
limits for the spin-independent scattering cross section are shown in
the usual normalization as a cross section per nucleon which is the
only practical method to compare results from experiments with
different target materials. What is also shown as a shaded region is
the detection rate expected for neutralinos in the minimal
superymmetric standard model (MSSM) without universal scalar mass
unification~\cite{Gondolo,Bottino}. 

\begin{figure}[b]
\centerline{\psfig{figure=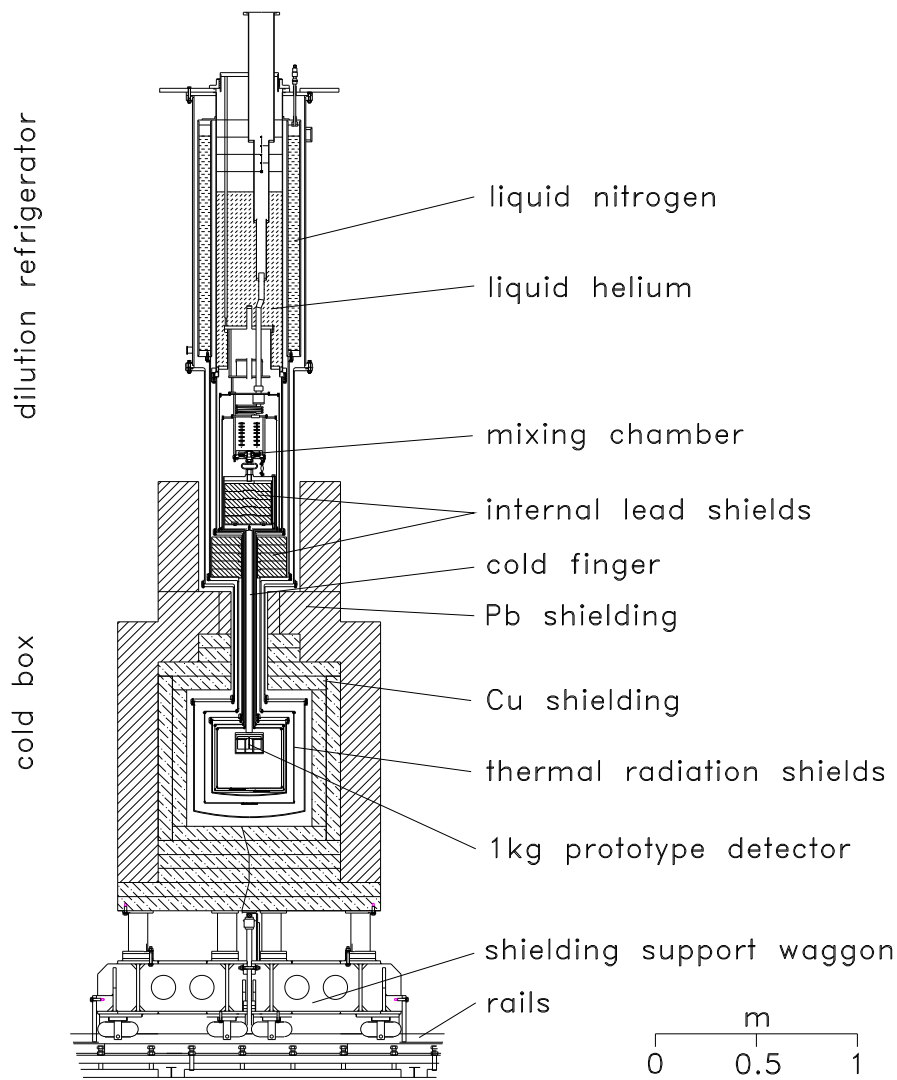,width=12.0cm}}
\caption{Schematic view of the experimental setup of the CRESST
experiment~\protect\cite{CRESST}, located in the Gran Sasso
underground laboratory near Rome (Italy), as an example for a
cryogenic dark matter experiment.
\label{fig:cresst}}
\end{figure}

Intruigingly, the current experiments already touch the parameter
space expected for supersymmetric particles and thus are in a position
where they begin to have a real chance of actually detecting dark
matter. One problem is, of course, how one could attribute a tentative
signal unambiguously to galactic WIMPs rather than some unidentified
radioactive background. One signature is the annual signal modulation
which arises because the Earth moves around the Sun while the Sun
orbits around the center of the galaxy. Therefore, the net speed of
the Earth relative to the galactic dark matter halo varies, causing a
modulation of the expected counting rate because of the modulation of
the effective WIMP velocity distribution seen by the detector. The
DAMA/NaI experiment has actually reported tentative evidence for such
a modulation~\cite{DAMA97} which would point to neutralinos just below
their previous exclusion range \cite{Bottino97}. At the present time
the significance of this signature is very low, and tentative signals
are bound to appear just below the previous exclusion ranges. Still,
the good news is that this tentative claim {\em could\/} be true in
the sense that one has reached the sensitivity necessary to find
supersymmetric dark matter.

In the near future large-scale cryogenic detectors will explore a vast
space of WIMP-nucleon cross-sections. The CRESST
experiment~\cite{CRESST} (Fig.~\ref{fig:cresst}) which is located in
the Gran Sasso underground laboratory aims at relatively low WIMP
masses with a sensitivity goal indicated by the dotted line in
Fig.~\ref{fig:limits}.  It uses sapphire crystals as targets and a
purely bolometric technique to measure the heat deposition by WIMP
collisions.  The CDMS experiment~\cite{CDMS} is currently located at a
shallow site at Stanford, but will eventually move to a deep site in
the Soudan mine. It uses germanium detectors and thus can discriminate
against backgrounds by measuring simultaneously the bolometric and
ionization signal. Unless WIMPs show up with relatively large
interaction cross sections it will be inevitable to use some form of
background discrimination to cover a reasonably large range of
supersymmetry-motivated interaction cross~sections.

There exist other ``indirect'' methods to search for galactic
WIMPs~\cite{JKG96}. They would annihilate with a certain rate in the
galactic halo, producing a potentially detectable background of
high-energy photons or antiprotons. Moreover, they interact with the
matter that makes up the Earth or Sun so that a small fraction of the
WIMPs traversing these bodies will lose enough energy to be trapped
and to build up at their centers. The WIMP annihilation would thus
produce high-energy neutrinos from the center of the Earth and from
the Sun which are detectable in neutrino telescopes. The existing
limits~\cite{tele-obs} already begin to touch the parameter range
relevant for supersymmetric dark matter~\cite{tele-theory}.  Put
another way, neutrino telescopes are already competitive with direct
experiments at searching for dark matter. It depends on details of the
supersymmetric models and parameters if direct search experiments or
neutrino telescopes have a better chance of finding dark matter
neutralinos. Roughly speaking, though, an ice Cherenkov detector like
AMANDA at the south pole~\cite{AMANDA} requires a $\rm km^3$ volume to
be competitive with the CDMS-Soudan search goal. It is to be expected
that AMANDA can actually be upgraded to this volume within the next
five years or so, providing neutrino astronomy with a good chance of
detecting the dark matter of our galaxy.

%%%%%%%%%%%%%%%%%%%%%%%%%%%%%%%%%%%%%%%%%%%%%%%%%%%%%%%%%%%%%%%%%%%%%%
%%%%%%%%%%%%%%%%%%%%%%%%%%%%%%%%%%%%%%%%%%%%%%%%%%%%%%%%%%%%%%%%%%%%%%

\subsection{Axions}

Axions are a particle dark-matter candidate sui generis in that they
are very weakly interacting very low-mass bosons and are yet a
candidate for cold dark matter, in apparent defiance of the
Tremaine-Gunn argument. However, contrary to neutrinos or WIMPs,
axions were never in thermal equilibrium in the early universe; they
appear in the form of highly occupied and thus essentially classical
oscillations of the axion field.

The existence of axions is motivated by the CP problem of QCD which
consists of the observed smallness of a possible neutron electric
dipole moment relative to a naive QCD expectation which would put it
at roughly the same magnitude as the neutron magnetic dipole
moment. Put another way, because of its nontrivial vacuum structure
QCD is expected to produce CP violating effects which are measured by
a parameter $\Theta$ which is an angular variable and thus can take on
any value between $0$ and $2\pi$.  The experimental limits on the
neutron electric dipole moment (a CP-violating quantity) reveal that
$\Theta\alt 10^{-9}$ while there is no a-priori reason why it should
not be of order unity.  The Peccei-Quinn solution~\cite{PecceiQuinn}
to this CP problem of the strong interaction (``strong CP problem'')
 is
based on re-interpreting $\Theta$ as a physical field by virtue of
$\Theta \to a(x)/f_a$, where $a(x)$ is the pseudoscalar axion field
while $f_a$ is an energy scale known as the Peccei-Quinn scale or
axion decay constant.  The main aspect of the Peccei-Quinn mechanism
is that the CP-violating Lagrangian produces a potential which drives
the axion-field to the CP-conserving position corresponding to
$\Theta=0$ so that CP violation is switched off by its own force.
This dynamical way of restoring CP conservation can be pictured in an
intruiging mechanical analogy devised by Sikivie~\cite{Sikivie96a}.

The unavoidable quantum excitation of the new field are the
axions~\cite{WW}.  Apart from model-dependent fine points, all of
their properties are fixed by the value of $f_a$; for reviews see
Refs.~\cite{axionreviews}. Phenomenologically, axions are closely
related to neutral pions; they mix with each other with an amplitude
of about $f_\pi/f_a$ where $f_\pi=93\,\rm MeV$ is the pion decay
constant.  Therefore, the axion mass and interactions follow roughly
by scaling the corresponding $\pi^0$ properties with $f_\pi/f_a$; for
example, $m_a f_a\approx m_\pi f_\pi$.  The axion couplings to photons
or nucleons is inversely proportional to $f_a$ and thus arbitrarily
small if $f_a$ is sufficiently large. Analogous to $\pi^0$ axions have
a two-photon coupling ${\cal L}_{a\gamma}=g_{a\gamma} {\bf E}\cdot{\bf
B}\,a$ where ${\bf E}$ and ${\bf B}$ are the electric and magnetic
field strengths. The coupling constant is $g_{a\gamma}=-\alpha/2\pi
f_a$ times a model-dependent factor of order unity.  Thus far axions
have not been detected in any laboratory experiment.  In addition,
their interaction strength can be constrained by demanding that they
do not carry away more energy from the interior of stars than is
compatible with astronomical observations~\cite{RaffeltBook}. The
limits on $f_a$ and $m_a$ thus obtained imply that axions must be very
light ($m_a\alt 10^{-2}\,\rm eV$) and very weakly interacting if they
exist at all.

In order to understand the cosmological evolution of axions note that
in concrete implementations of the Peccei-Quinn mechanism the axion
field is interpreted as the phase of a new Higgs field $\Phi(x)$ which
undergoes the spontaneous breakdown of a chiral $U(1)$ symmetry, the
Peccei-Quinn symmetry. The potential which causes the symmetry
breaking is a ``Mexican hat'' with a vacuum expectation value of the
ground state somewhere in the rim of the hat.  The axion field is the
angular degree of freedom, i.e.\ the axion is the Nambu-Goldstone
boson of the spontaneously broken Peccei-Quinn symmetry.  In the very
early universe when the temperature falls below $f_a$ the Peccei-Quinn
symmetry breaks down, meaning that $\Phi(x)$ needs to find its minimum
somewhere in the rim of the Mexican hat, i.e.~it needs to choose one
value for the axion field $a(x)$ or equivalently for the CP-violating
QCD parameter $\Theta$.  Later at a temperature $T=\Lambda_{\rm
QCD}\approx200\,\rm MeV$ the QCD phase transition occurs which implies
that the potential for the axion field is switched on, driving it to
the CP-conserving minimum. One may equally say that at the QCD phase
transition the Peccei-Quinn symmetry is explicitly broken, that the
Mexican hat tilts, or that the axion mass turns on. The axion no
longer is a (strictly massless) Nambu-Goldstone boson, it has become a
(low-mass) pseudo Nambu-Goldstone boson.

The initial ``misalignment'' $\Theta_i$ of the axion field relative to
the CP-conserving minimum of the QCD-induced potential sets the axion
field into motion and thus excites coherent
oscillations~\cite{Misalignment}. They correspond to an axionic mass
density of the present-day universe of about
\begin{equation}
\Omega_a h^2\approx1.9\times 4^{\pm1} 
(\mu{\rm eV}/m_a)^{1.175}\,
\Theta_i^2\,F(\Theta_i).
\end{equation} 
The stated range reflects recognized uncertainties of the cosmic
conditions at the QCD phase transition and uncertainties in the
calculation of the temperature-dependent axion mass. The function
$F(\Theta)$ with $F(0)=1$ encapsules anharmonic corrections to the
axion potential.  If $\Theta_i$ is of order unity, axions with
$m_a={\cal O}(1\,\mu{\rm eV})$ provide roughly the cosmic closure
density. The equivalent Peccei-Quinn scale $f_a={\cal O}(10^{12}\,{\rm
GeV})$ is far below the GUT scale so that one may speculate that
cosmic inflation, if it occurred at all, did not occur after the PQ
phase transition.  

If it did not occur at all, or if it did occur before the PQ
transition with $T_{\rm reheat}>f_a$, the axion field will start with
a different $\Theta_i$ in each region which is causally connected at
$T\approx f_a$. Then one has to average over all regions to obtain the
present-day axion density.  More importantly, because axions are the
Nambu-Goldstone mode of a complex Higgs field after the spontaneous
breakdown of a global U(1) symmetry, cosmic axion strings will form by
the Kibble mechanism~\cite{Davis}.  The motion of these global strings
is damped primarily by the emission of axions rather than
gravitational waves. At the QCD phase transition the U(1) symmetry is
explicitly broken (axions acquire a mass) causing domain walls bounded
by strings to form which get sliced up by the interaction with
strings. The whole string and domain-wall system will quickly decay
into axions. This complicated sequence of events leads to the
production of the dominant contribution of cosmic axions where most of
them are produced near the QCD transition. After they acquire a mass
they are nonrelativistic or mildly relativistic so that they are
quickly redshifted to nonrelativistic velocities.  The proper
treatment of axion radiation by global strings is difficult and has
been partly controversial. However, taking account of all recognized
uncertainties one arrives at a plausible range for the mass of
dark-matter axions between a few $\mu \rm eV$ and a few~$\rm meV$.

The axions produced by strings or the misalignment mechanism were
never in thermal equilibrium; the field modes are highly occupied,
forming something like a Bose-Einstein condensate. Axions are
nonrelativistic almost from the start and thus form cold dark matter,
in spite of their small mass.  If the axion interaction were
sufficiently strong ($f_a\alt10^8\,{\rm GeV}$) they would have come
into thermal equilibrium before the QCD phase transition, leading to
an axion background in analogy to the one expected for neutrinos
\cite{TurnerI}. However, this parameter range is excluded by the
astrophysical arguments which imply that axions interact so weakly
that they have never come into thermal equilibrium in the early
universe. They cannot provide a hot dark matter component.

\begin{figure}[ht]
\centerline{\psfig{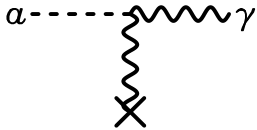}}
\caption{Primakoff conversion of axions into photons in the presence
of an external electromagnetic field.
\label{fig:ax0}}
\end{figure}

If axions are the galactic dark matter one can search for them in the
laboratory. The detection principle is analogous to the Primakoff
effect for neutral pions which can convert into photons in the
presence of an external electromagnetic field due to their two-photon
vertex (Fig.~\ref{fig:ax0}). Dark matter axions would have a mass in
the $\rm \mu eV$ to meV range. Because they are bound to the galaxy
their velocity dispersion is of order the galactic virial velocity of
around $10^{-3} c$ so that their kinetic energy is exceedingly small
relative to their rest mass. Noting that a frequency of $1\,\rm GHz$
corresponds to $4\,\rm \mu eV$ the Primakoff conversion produces
microwaves. Because the galactic axions are nonrelativistic
while the resulting photons are massless the conversion involves a
huge momentum mismatch which can be overcome by looking for the
appearance of excitations of a microwave cavity rather than for free
photons.

An axion search experiment thus consists of a high-$Q$ microwave
resonator placed in a strong external magnetic field (``axion
haloscope'' \cite{Sikivie}).  The microwave power output of such a
cavity detector on resonance is~\cite{Sikivie,Krauss}
\begin{equation}
P\approx0.4\times10^{-22}\,{\rm Watts}
\left(\frac{V}{0.2\,\rm m^3}\right)
\left(\frac{B}{7.7\,\rm Tesla}\right)^2
\left(\frac{C}{0.65}\right)
\left(\frac{Q}{10^5}\right)
\left(\frac{\rho_a}{300\,\rm MeV\,cm^{-3}}\right)
\left(\frac{m_a}{1\,\rm \mu eV}\right),
\end{equation}
where $V$ is the cavity volume, $B$ the applied magnetic field
strength, $C$ a mode-dependent form factor which is largest for the
fundamental T$_{010}$ mode, $Q$ the loaded quality factor, and
$\rho_a$ the local galactic axion density. This is indeed a weak
microwave signal! If $m_a$ were known it would be easy to detect
galactic axions with this method. One may verify or reject a tentative
signal by varying, for example, the applied magnetic field
strength. Therefore, contrary to the WIMP experiments it would be hard
to mistake a background signal for a dark-matter signature. The
problem is, of course, that $m_a$ is not known so that one needs a
tunable cavity, stepping its resonance through as large a frequency
range as possible and to look for the appearance of microwave power
beyond thermal and amplifier noise.

\begin{figure}[ht]
\centerline{\psfig{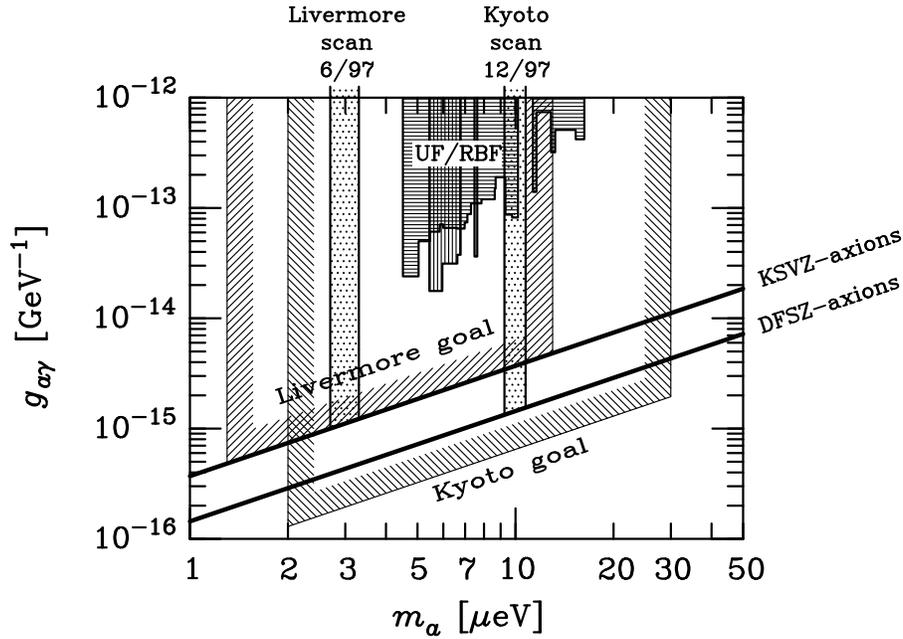}}
\caption{Current limits on galactic dark matter axions from the
  University of Florida (UF) \protect\cite{UFexperiment} and the
  Rochester-Brookhaven-Fermilab (RBF) \protect\cite{RBFexperiment}
  search experiments and search goals of the current
  Livermore~\protect\cite{Livermore} and Kyoto~\protect\cite{Kyoto}
  experiments. It was assumed that the local galactic axion density is
  $300\,\rm MeV\,cm^{-3}$.
\label{fig:ax2}}
\end{figure}

Two pilot experiments of this sort~\cite{UFexperiment,RBFexperiment}
have excluded the range of axion masses and coupling strengths
indicated in Fig.~\ref{fig:ax2}. Evidently, for a standard local halo
density of about $300\,\rm MeV\,cm^{-3}$ they were not sensitive
enough to reach realistic axion models. Two current experiments with
much larger cavity volumes, however, have the requisite sensitivity;
their search goals are depicted in Fig.~\ref{fig:ax2}. In its current
setup, the Livermore experiment~\cite{Livermore} uses conventional
microwave amplifiers which limit the useful cavity temperature to
about $1.4\,\rm K$. The Kyoto experiment CARRACK~\cite{Kyoto}, on the
other hand uses a completely novel detection technique based on the
excitation of a beam of Rydberg atoms which passes through the
cavity. This is essentially a counting method for microwaves which
does not require a (noisy) amplifier so that one can go to much lower
physical cavity temperatures. This enhances the sensitivity and also
allows one to use smaller cavity volumes and thus to search for larger
axion masses.

In summary, the second generation axion experiments have reached a
sensitivity where they may well turn up axion dark matter during their
expected running time of a few years. If they fail to find axions it
would be extremely important to extend the experimental search into a
regime of larger masses toward the meV scale.

%%%%%%%%%%%%%%%%%%%%%%%%%%%%%%%%%%%%%%%%%%%%%%%%%%%%%%%%%%%%%%%%%%%%%%
%%%%%%%%%%%%%%%%%%%%%%%%%%%%%%%%%%%%%%%%%%%%%%%%%%%%%%%%%%%%%%%%%%%%%%

\subsection{Primordial Black Holes}

Stellar-remnant black holes collapse from baryonic material and are
thus probably excluded as dark-matter
candidates~\cite{Carr}. Primordial black holes which form before
big-bang nucleosynthesis, on the other hand, are perfect cold dark
matter candidates; in a sense they are just particularly fat
WIMPs. The microlensing observations of apparent MACHOs as a
significant dark-matter component of the galactic halo has revived the
interest in these objects. The main objection against them is the lack
of a plausible mechanism for making them in the early universe, even
though there have been some intruiging recent
suggestions~\cite{PBH}. In any event, as long as particle dark matter
remains undiscovered the option of primordial black holes as a CDM
candidate should not be forgotten.

%%%%%%%%%%%%%%%%%%%%%%%%%%%%%%%%%%%%%%%%%%%%%%%%%%%%%%%%%%%%%%%%%%%%%%
%%%%%%%%%%%%%%%%%%%%%%%%%%%%%%%%%%%%%%%%%%%%%%%%%%%%%%%%%%%%%%%%%%%%%%

\subsection{Modified Gravity}

The hypothesis of particle dark matter requires nontrivial and perhaps
bewildering extensions of the particle-physics standard model.  As
long as the nature of dark matter has not been positively identified
it may seem no more radical to try to modify general relativity such
that there is no need for dark matter.  It has sometimes been argued
that the hypothesis of dark matter is just a parametrization of our
ignorance of the physical laws which apply on large astrophysical
scales where no independent test of the validity of general relativity
exists that would not involve the hypothesis of dark matter.

In one phenomenological approach known as MOND for Modified Newtonian
Dynamics~\cite{Milgrom94} gravitational accelerations $a$ below a
certain limit $a_0$ are given by $a^2/a_0=G_{\rm N} M/r^2$, where
$G_{\rm N}$ is Newton's constant. With $a_0\approx10^{-8}\,\rm
cm\,s^{-2}$ this approach is surprisingly successful at explaining a
broad range of dark-matter phenomena related to dwarf galaxies, spiral
galaxies, and galaxy clusters~\cite{Milgrom94,Milgrom95}.
Unfortunately, MOND lacks a relativistic formulation so that it cannot
be applied on cosmological scales.

One covariant alternative to general relativity is a conformally
invariant fourth-order theory~\cite{Mannheim95}.  In the
nonrelativistic regime it leads to a linear potential in addition to
the Newtonian $r^{-1}$ term. It explains at least some of the galactic
and cluster dark-matter problems.

Before modifications of general relativity can be taken seriously they
must pass relativistic tests. An important case are galaxy clusters
where large amounts of dark matter are indicated by nonrelativistic
methods (virial theorem) as well as by relativistic indicators
(gravitational lensing, notably giant arcs). Because virial and
lensing masses seem to agree well in several cases, scalar-tensor
extensions of general relativity are in big trouble, if not ruled out
entirely~\cite{Bekenstein94}. One way out could be a certain
preferred-frame theory which can reproduce the MOND phenomenology as
well as the lensing effects~\cite{Sanders97}.

Apparently, no serious attempt has been made to discuss truly
cosmological phenomena such as structure formation and cosmic
microwave background distortions in the framework of alternative
theories of gravity. At the present time it is not known if a
covariant theory of gravity exists that can explain the dark-matter
problems on all scales. However, as long as the nature of dark matter
has not been identified one should keep an open mind to such
possibilities!

%%%%%%%%%%%%%%%%%%%%%%%%%%%%%%%%%%%%%%%%%%%%%%%%%%%%%%%%%%%%%%%%%%%%%%
%% Section V %%%%%%%%%%%%%%%%%%%%%%%%%%%%%%%%%%%%%%%%%%%%%%%%%%%%%%%%%
%%%%%%%%%%%%%%%%%%%%%%%%%%%%%%%%%%%%%%%%%%%%%%%%%%%%%%%%%%%%%%%%%%%%%%

\section{CONCLUSION}

There is now little doubt that the dynamics of the universe on
galactic scales and above is dominated by dark matter which almost
certainly is not in the form of objects which are familiar to us. Much
of the evidence points in the direction of a cosmic background
of new weakly interacting particles, with neutralinos, axions, and
neutrinos the favored options because they are well motivated by
particle-physics theory for reasons other than pleasing the
astronomers. At the present time the existence of dark matter is
perhaps the strongest empirical evidence for particle physics beyond
the standard model.

Over the past decade one has become used to the idea that most of the
stuff in the universe consists of nonbaryonic matter.  Yet this
remains a radical conjecture which has often been likened to the
Copernican revolution when Earth and with it Man was moved from the
center of creation to some unspectacular average position.  Probably
the next big step in the Second Copernican Revolution will be the
final deciphering of the ``Cosmic Rosetta Stone'' in the form of
precision measurements of the angular temperature fluctuations of the
cosmic microwave background which will confirm or refute the apparent
discrepancy between the baryon content of the universe and its
dynamical mass density. Even then, however, this second revolution
will not be complete without a direct and positive identification of
the dark matter particles or objects. Therefore, the search
experiments for galactic dark matter as well as the laboratory
searches for supersymmetric particles and neutrino masses are among
the most important scientific enterprises in our attempt to understand
the universe.

\begin{figure}[t]
\centerline{\psfig{figure=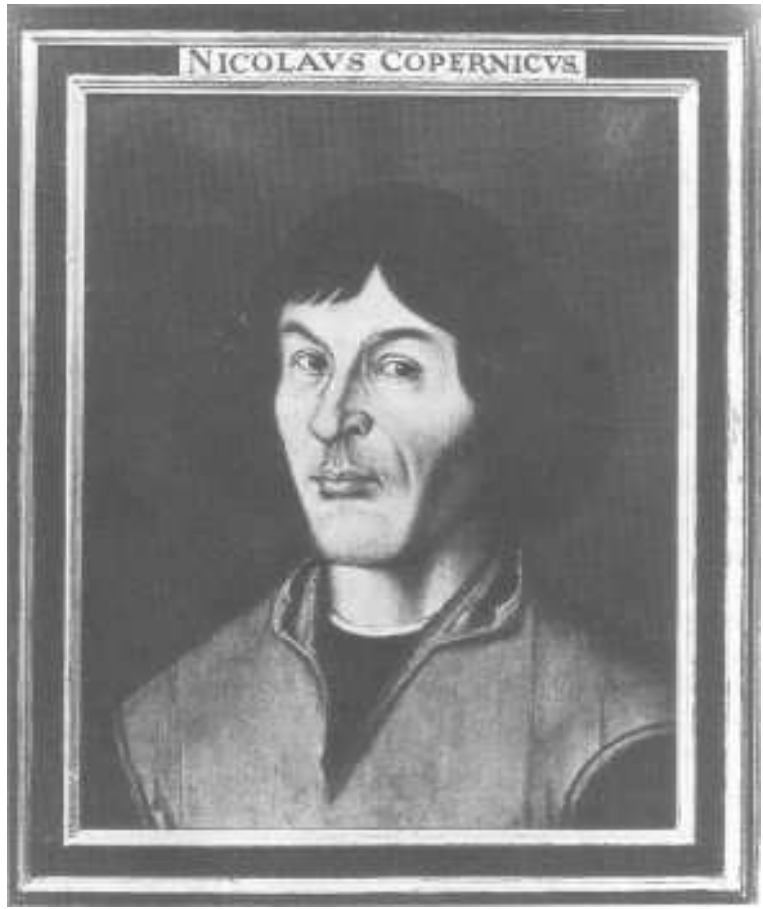,width=10.0 cm}}
\caption{Nicolaus Copernicus (1473$-$1543).
\label{fig:nico}}
\end{figure}

%%%%%%%%%%%%%%%%%%%%%%%%%%%%%%%%%%%%%%%%%%%%%%%%%%%%%%%%%%%%%%%%%%%%%%
%% Acknowledgments %%%%%%%%%%%%%%%%%%%%%%%%%%%%%%%%%%%%%%%%%%%%%%%%%%%
%%%%%%%%%%%%%%%%%%%%%%%%%%%%%%%%%%%%%%%%%%%%%%%%%%%%%%%%%%%%%%%%%%%%%%

\subsection*{Acknowledgments}

This work was supported, in part, by the European Union under contract
No.\ CHRX-CT93-0120 and by the Deutsche Forschungsgemeinschaft under
grant No.\ SFB~375. 

%%%%%%%%%%%%%%%%%%%%%%%%%%%%%%%%%%%%%%%%%%%%%%%%%%%%%%%%%%%%%%%%%%%%%%
%% References %%%%%%%%%%%%%%%%%%%%%%%%%%%%%%%%%%%%%%%%%%%%%%%%%%%%%%%%
%%%%%%%%%%%%%%%%%%%%%%%%%%%%%%%%%%%%%%%%%%%%%%%%%%%%%%%%%%%%%%%%%%%%%%

\end{document}